\pdfoutput=1

\documentclass[american,preprint,preprintnumbers,prb,showpacs,superscriptaddress]{revtex4-1}

\usepackage{amsmath}
\usepackage{graphicx}
\usepackage{amssymb}
\usepackage{times}
\usepackage{color}
\usepackage{float}
\usepackage{subcaption}
\definecolor{lr}{rgb}{1.0,0.3,0.3}
\definecolor{dg}{rgb}{0.0,0.5,0.0}
 


\begin{document}

\title{First principles predictions of magneto-optical data for semiconductor defects: the case of divacancy defects in 4H-SiC}

\author{Joel Davidsson}
\email{joel.davidsson@liu.se}
\affiliation{Department of Physics, Chemistry and Biology, Link\"oping
  University, SE-581 83 Link\"oping, Sweden}

\author{Viktor Iv\'ady}
\affiliation{Department of Physics, Chemistry and Biology, Link\"oping
  University, SE-581 83 Link\"oping, Sweden}
\affiliation{Wigner Research Centre for Physics, Hungarian Academy of Sciences,
  PO Box 49, H-1525, Budapest, Hungary}

\author{Rickard Armiento}
\affiliation{Department of Physics, Chemistry and Biology, Link\"oping
  University, SE-581 83 Link\"oping, Sweden}

\author{N.T.\ Son}
\affiliation{Department of Physics, Chemistry and Biology, Link\"oping
  University, SE-581 83 Link\"oping, Sweden}

\author{Adam Gali} 
\affiliation{Wigner Research Centre for Physics, Hungarian Academy of Sciences,
  PO Box 49, H-1525, Budapest, Hungary}
\affiliation{Department of Atomic Physics, Budapest University of
  Technology and Economics, Budafoki \'ut 8., H-1111 Budapest,
  Hungary}

\author{Igor A. Abrikosov}
\affiliation{Department of Physics, Chemistry and Biology, Link\"oping
  University, SE-581 83 Link\"oping, Sweden}
\affiliation{Materials Modeling and Development Laboratory, National University of Science and Technology `MISIS', 119049 Moscow, Russia}


\begin{abstract}
Study and design of magneto-optically active single point defects in semiconductors are rapidly growing fields due to their potential in quantum bit and single photon emitter applications. Detailed understanding of the properties of candidate defects is essential for these applications, and requires the identification of the defects microscopic configuration and electronic structure. Multi-component semiconductors often host two or more non-equivalent configurations of point defects. These configurations generally exhibit similar electronic structure and basic functionalities, however, they differ in details that are of great importance whenever single defect applications are considered. Identification of non-equivalent configurations of point defects is thus essential for successful single defect manipulation and application. A promising way to identify defects is via comparison of experimental measurements and results of first-principle calculations. We investigate a possibility to produce accurate \emph{ab initio} data for zero-phonon lines and hyperfine coupling parameters that are required for systematic quantum bit search. We focus on properties relevant for the possible use of the divacancy defect in quantum bits in 4H-SiC. We provide a decisive identification of divacancy configurations in 4H-SiC and clarify differences in prior predictions of 4H-SiC divacancy zero-phonon photoluminescence lines.

\end{abstract}

\maketitle


\section{Introduction}

The physics of semiconductor point defects is of outstanding importance for controlling their optical and electrical properties\cite{BookPointDefect1,BookPointDefect2}.The study of point defect properties is a field of much active interest due to recent discoveries of numerous magnetically and optically active defect centers that can act as a single photon source \cite{Childress:PRL2006,Aharonovich:NL2009,Kolesov2012,NatMat14,Aharonovich2016} or a quantum bit (qubit) \cite{Jelezko:PSS2006,Hanson:Nature2008,Awschalom2013}. So far, the most thoroughly investigated point defect for use in qubits are the NV-center in diamond\cite{duPreez:1965,Balasubramanian:NatMat2009,Awschalom:Nature2010,Robledo:Nature2011},  phosphor in silicon\cite{Morton2008,Morello2010,Pla2012} and divacancy\cite{Koehl11,Falk2013,Christle2014} in silicon carbide (SiC). Furthermore, numerous other centers in various semiconducting host materials are proposed as potential magneto-optical centers, such as silicon-vacancy and germanium-vacancy centers in diamond\cite{SipahigilPRL2014,Iwasaki2015}, silicon vacancy in SiC\cite{Soltamov12,NatPhys14}, carbon anti-site vacancy pair in SiC\cite{Szasz2015}, Ce$^{3+}$ and Pr$^{3+}$ ions in yttrium aluminium garnet\cite{KolesovPRL2013,Kolesov2012}, Eu and Nd$^{3+}$ion in yttrium orthosilicate\cite{LongdellPRL2014,Clausen2011}, Nd$^{3+}$  yttrium orthovanadate\cite{DeRiedmatten2008}, defect spins in aluminum nitride\cite{Seo2016}, etc.

To manipulate these centers on single defect level and to reconstruct their Hamiltonian, it is essential to identify the microscopic structure, electronic structure, and spin-configuration of the center. State-of-the-art experimental teqchniques used in experimental point defect investigation are for instance, photoluminescence (PL) or absorption spectroscopy, electron spin resonance (ESR), deep-level transient spectroscopy (DLTS), and Raman spectroscopy, that probe different characteristics of the centers. Gathering all the available information about a considered center can provide an appropriate working model. However, there are numerous unidentified defect centers in most of the commercially available semiconductors\cite{magnusson2005}. 

In semiconductors where multiple non-equivalent sites exist in the primitive cells, each point defect can have several different configurations. These distinguishable configurations exhibit different properties and thus different applicability in qubit and single photon emitter applications. The identification of such non-equivalent configurations is particularly challenging. For the non-equivalent configurations of divacancy related qubits in 4H-SiC two contradictory identifications have been presented, which rely on either the calculated zero-phonon photoluminescence (ZPL) lines\cite{Gordon2015} or the zero-field splitting parameter (ZFS)\cite{Falk2014}. Furthermore, recently more divacancy related centers were reported than the possible number of non-equivalent divacancy configurations in SiC\cite{Koehl11,Falk2013}, which makes the identification even more puzzling.

Identification and characterization of point defects are greatly facilitated by first principles theory. In supercell or cluster models, a small part of the material that embeds a single point defect is directly modeled in electronic structure calculations. This way many properties of the defects, such as spectral properties, charge transition levels, ESR parameters can be obtained, etc. In the literature, one can find different strategies how these quantities can be obtained in first principles calculations\cite{Kohan2000,Janotti2007,LanyZunger08,Deak2010,FreysoldtRMP2014}. However, when comparing computational results to experiments, special care must be taken to account for numerical uncertainty and limitations in the theoretical methods\cite{FreysoldtRMP2014}. So far there has been limited discussion about how to generally achieve an accuracy sufficient for identification of non-equivalent defects. In the present paper, we address this issue.

We assess the accuracy of first principle calculations of ZPL and hyperfine interaction parameters to create guidelines for theoretical point defect calculations that allow non-equivalent defect configurations to be identified. In particular, we consider the divacancy defect in 4H-SiC and consistently identify PL1-PL4 room temperature qubits by comparing convergent magneto-optical date with the experiment. This defect has been studied with \emph{ab initio} calculations before, but the present study helps clarifying previous results that (as discussed above) have not been fully consistent\cite{Gordon2015,Falk2014}. However, a main purpose of the present investigation is also to identify a scheme capable of reliably generating data via high-throughput calculations\cite{ceder2013supercomputers,curtarolo2013high} useful for identification of, essentially, any previously unknown point defect. For this intended use, it is imperative to identify methods that reliably produce sufficiently accurate results, but also take minimal computational effort.

The rest of this paper is organized as follows. Section~II\ describes the basic properties of SiC and divacancy point defect in 4H polytype of SiC. In Section~III\ gives details on the first principle methods used in this work. Section~IV\ presents the results and discussion of our first principles point defect calculations. In section~V, we demonstrate how to use our results to identify divacancy configurations in 4H-SiC. Finally, section~VI\ summarizes our findings.

\section{Divacancy in SiC}

\begin{figure}[h!]
	\includegraphics[width=0.6\columnwidth]{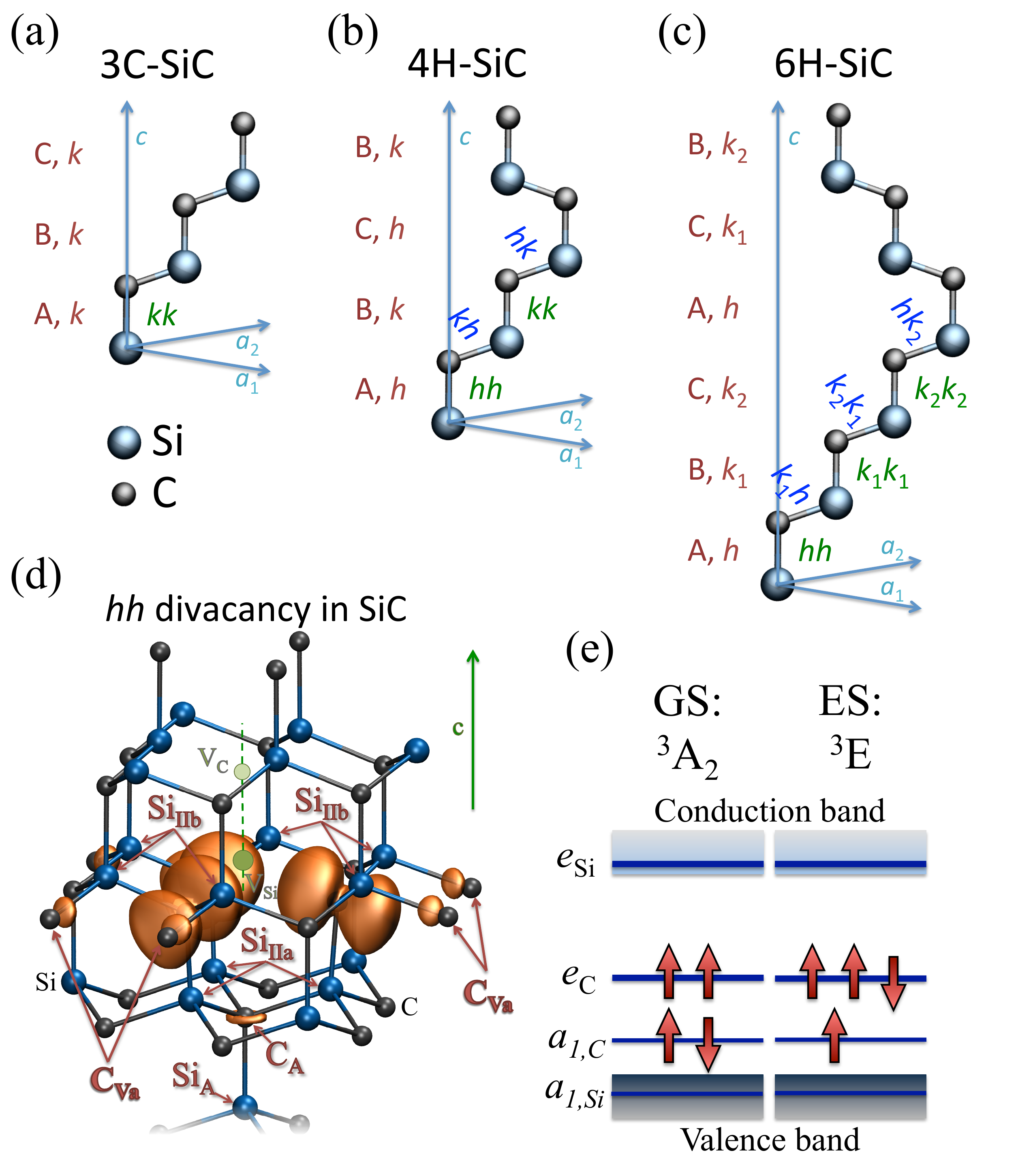}
	\caption{Structure and configuration of defects in SiC. Fig (a-c) show the primitive cells of 3C, 4H, and 6H-SiC. Red upper-case letters show the stacking of Si-C double layers, while lower-case letters shows whether the double layers and their immediate surroundings follow a cubic like ($k$) or in a hexagonal-like ($h$) stacking order. Green lower-case letter pairs show variants of a pair defect. Figure (d) depicts $hh$ divacancy configuration in 4H-SiC, the spatial distribution of the spin density (orange lobes), and the important neighboring Si and C sites that can give rise to notable hyperfine interaction. Figure (e) schematically shows the occupancy of single-particle orbitals in the ground and excited states of the divacancy.} 
	\label{fig:divac}  
\end{figure}

SiC is a polytypic semiconductor with more than 250 polytypes synthesized. The most commonly used forms are 3C, 4H, and 6H-SiC. The 3C polytype, shown in Fig.~\ref{fig:divac}(a), has cubic symmetry with a single C and Si atom in the primitive cell. The 4H polytype, in Fig.~\ref{fig:divac}(b), has hexagonal symmetry and 8 atoms in the primitive cell of which 2 are non-equivalent for both Si and C (see Fig.~\ref{fig:divac}(a)). The 6H polytype is also of hexagonal symmetry, has 12 atoms in the primitive cell, and 3 are non-equivalent for both Si and C (see Fig.~\ref{fig:divac}(c)). Hence, a single site defect in 4H has 2 distinguishable configurations, and 3 in 6H. A pair defect then has 4 and 6 configurations, respectively.  The non-equivalent sites in 4H and 6H-SiC are refered to as $h$ and $k$ (4H), and $h$, $k_1$, $k_2$ (6H). Here, $h$ refers to a site in an hexagonal-like environment, and $k$ to a cubic-like environment. In this paper, we focus on the four possible configurations of divacancy in 4H-SiC; $hh$, $kk$, $hk$, and $kh$, where the V$_{\text{Si}}-$V$_{\text{C}}$ notation is used. For two of these configurations, $hh$ and $kk$, the  V$_{\text{Si}} - $V$_{\text{C}}$ axis of the defect is parallel to the hexagonal axis of 4H-SiC and possess $C_{3v}$ point group symmetry. The other two, $hk$ and $kh$, have lower $C_{1h}$ symmetry.  $hh$ and $kk$ configurations are often called as axial configurations, while $hk$ and $kh$ as basal configurations.  

Due to the $C_{3v}$ symmetry in $hh$ and $kk$, the dangling bonds of silicon and carbon vacancies form two fully symmetric $a_1$ and two double degenerate $e$ states, which are occupied by 6 electrons\cite{Gali10}. The $a_1$ states are fully occupied, with the one localized on the silicon dangling bonds falling into the valence band, while the other one is localized on the carbon atoms and appears in the band gap of 4H-SiC near the conduction band edge. The $e$ state is localized on the carbon dangling bonds and is located in the middle of the band gap and occupied by two electrons with parallel spin in the spin-1 ground state the divacancy. The other empty $e$ state falls into the conduction band, as shown in Fig.~\ref{fig:divac}(e).  In the case of basal configurations, the low symmetry crystal field splits the $e$ states into $a'$ and $a''$ and transforms $a_1$ to $a'$.

Due to the spin-1 ground state and localized nature of the defect states, a strong dipole-dipole interaction can be observed between the unpaired electrons which causes a splitting of the spin sublevels even at zero magnetic field. For the divacancy defect, this zero-field-splitting is approximately 1.3~GHz. In SiC there are two intrinsic paramagnetic nuclei, the spin-1/2 $^{13}$C with 1.07\% natural abundance and the spin-1/2 $^{29}$Si with 4.68\% natural abundance, which can interact with the spin of divacancy and cause hyperfine structure in ESR spectrum. The spin density and important nuclei sites which yields resolvable hyperfine splitting of $10 - 100$~Mhz are shown in Fig.~\ref{fig:divac}(d).

In the single particle picture, the optically excited state of lowest energy can be constructed by a spin conserving promotion of the electron from the higher $a_1$ state to the $e$ state, see Fig.~\ref{fig:divac}(e). Due to the partial occupancy of the $e_{\text{C}}$ state, the excited state is Jahn-Teller unstable, which causes spontaneous distortion of the atomic configurations of axial divacancies. In the many particle picture, six multiplets form in the triplet excited state that split according to the spin-spin and spin-orbit interactions. The divacancy in 4H and 6H-SiC has an electronic configuration similar to that of the divacancy and NV-center in diamond \cite{duPreez:1965,Doherty2013,Gali10,Koehl11}, thus the many particle picture derived for the NV-center\cite{Maze:2010,Doherty2011,Doherty2013} can be applied for the divacancy too.

The experimental divacancy related ZPL lines are named the UD-2 group\cite{magnusson2005} and PL1-4 lines\cite{Koehl11,Falk2013}, the electron spin resonance lines as P6/P7 centers\cite{Baranov2005,Son2006}. 

\section{Methodology}
The ZPL line is the energy difference between the ground state and excited state. These states can be seen in Fig.~\ref{fig:divac}(e). The energy is obtained by using Kohn-Sham (KS) density functional theory\cite{Hohenberg64,Kohn65} (DFT). In the excited state calculation, constrained occupation DFT scheme\cite{Gali:PRL2009} is applied, and accordingly a KS particle is promoted from the $a_1$ state to the empty $e$ state in the minority spin channel and self-consistent energy minimization, including geometry relaxation, is carried out. In \emph{absolute values}, one cannot expect better than 100 meV accuracy from this scheme, due to the single Slater determinant description of the excited state. This uncertainty of the theoretical method is an order of magnitude larger than the accuracy requirement of non-equivalent configuration identification (10 meV, which is the typical difference of ZPL energies for the divacancy defect). On the other hand, as the crystal field potential differs qualitatively only from the second neighbor shell and the defect state localization is decaying exponentially in this region, the electronic structure of the non-equivalent configurations can be considered as nearly identical. Hence, ZPL energy differences follow from the potential perturbations acting on the defect orbitals. Therefore, to identify the non-equivalent configurations, our DFT calculations must capture those effects that are caused by a perturbing crystal field potential. As the potential has a direct effect on the density and energy, such identification is likely to be possible through DFT ZPL energy calculations. In other words, for \emph{relative differences} of the ZPL energies, one may expect better than 100 meV in constrained occupation DFT calculations.  In the following, this topic is investigated in details by assessing technical and theoretical limitations of ZPL energy calculations. 

We apply three exchange-correlation functionals in our calculations; the semi-local functionals of Perdew, Erzenerhof, and Burke (PBE)\cite{PBE} and of Armiento and Mattsson (AM05)\cite{AM05}; and the screened hybrid functional of Heyd, Scuseria, and Ernzerhof (HSE06)\cite{HSE03,HSE06}. The hybrid functional is computationally much more expensive than the semi-local functionals, however, the band gap of semiconductors are closer to experiment \cite{HSE05} and accurate results in hyperfine field\cite{Szasz2013} as well as in zero-phonon line calculations\cite{Gali:PRL2009,Deak2010}.  All functionals are computed using the PBE pseudopotential labeled 05jan2001 for C and 08april2002 for Si. 

The recommended procedure for the HSE06 hybrid functional is to start from a semi-local density, hence the following scheme is introduced. The ground state is converged by first running a ground state PBE calculation then ground state HSE06 calculation. For the excited state, first, an excited state PBE calculation is executed. Then, a single self-consistent ground state HSE06 calculation is performed to obtain a good starting wavefunction for the final HSE06 excited state calculation.

In practice, we employ the Vienna Ab initio Simulation Package (VASP) \cite{VASP,VASP2}, which uses the plane wave basis set and the projector augmented wave (PAW) \cite{PAW,Kresse99} method to describe the KS states and handle the effects of the core electrons. Since we need highly accurate results, we use comparatively high settings for those convergence parameters that are not further discussed in this study. The stopping criterion for the self-consistent field calculations and for the structural minimization are $10^{-6}$ eV respectively $10^{-5}$ eV (energy difference) for the PBE functional. For HSE06 functional, the settings are instead $10^{-4}$ eV and $10^{-2}$ eV/A (force difference). The grid for the Fast Fourier transformation (FFT) is set to twice the largest wave vector in order to avoid wrap around errors. For the HSE06 functional, the FFT grid for the exact exchange is set to the largest wave vector. This produces some noises in the forces but good energies. The above-described settings ensure a numerical accuracy in the order of 1 meV for the calculated total energies. A Monkhorst-Pack\cite{monkhorst1976special} k-point grid is used for those calculations that use more k-points than the gamma point. In the rest this work, unit cell atom counts always refer to the number of atoms in a pristine supercell, if not otherwise specfied.

The zero-point energy shift of the ZPL energies due to the different vibrational properties of the ground and excited states is assumed to be small enough to not interfere with our conclusions. It is neglected in the present study. 

For hyperfine field calculations\cite{Szasz2013}, we use the implementation included in VASP, which gives the hyperfine tensor that describes the interaction between nuclear spin and electronic spin. This interaction produces a small splitting in energy levels which can be measured in experiments.

For zero-field-splitting, we follow the method of Ref.~\cite{Ivady2014} that requires Kohn-Sham wavefunctions as obtained by VASP DFT calculations. Here, we use the plane wave part of PBE wavefunctions to calculate the zero-field splitting tensor, the one center contributions from the PAW potentials are neglected. This approximation give an error of a few perecent in the calculated values.\cite{Ivady2014}

\section{Accurate point defect calculations}
In this section, results from zero-phonon line energy, hyperfine field, and stress calculations are presented.

\subsubsection{Supercell size}
\begin{figure}[h!]
	\includegraphics[width=0.7\columnwidth]{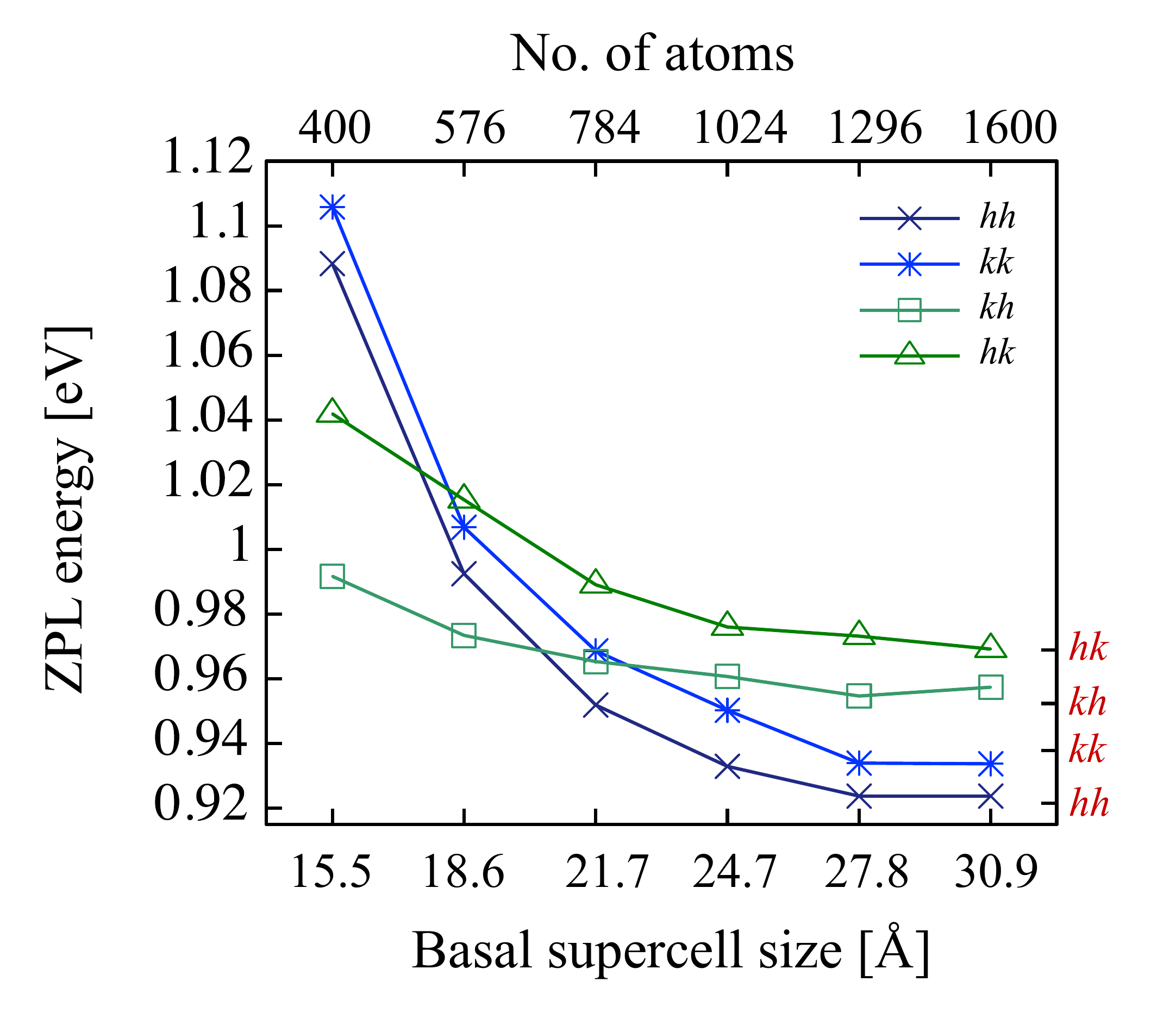}
	\caption{Supercell size convergence of ZPL energies of divacancy configurations in 4H-SiC calculated with the PBE functional and a $\Gamma$-point only k-point grid. Scaling is carried out in hexagonal supercell only in the basal direction. The $c$-axis size of the supercell is fixed at 20.25~\AA. The horizontal axis shows both the number of the atoms in the supercells and the basal plane lateral size of the supercell are provided. The right axis shows the converged ZPL for the 2400 supercell.} 
	\label{fig:size}  
\end{figure}

First, we demonstrate the convergence of ZPL energies on supercell size. We begin by using supercells that retain the hexagonal symmetry of the primitive cell, the PBE functional, and $\Gamma$-point sampling of the Brillouin zone (BZ). We fixed the $c$-axis size of the supercell to 20.25~\AA (twice of the primitive cell in that direction) and varied the lateral size of the cell in the basal directions. Figure~\ref{fig:size} shows how the ZPL energies converge with increasing supercell sizes. With a supercell of $10 \times 10$ copies of the primitive cell in the basal directions (ca 31~\AA \ in the $a1$- and $a2$-axis) the ZPL energies appear to converge to within 1 meV. Using this converged distance $\approx 30$~\AA\ in the $c$-direction as well gives a supercell ($10 \times 10 \times 3$) consisting of 2400 atoms. We will use the converged values at this size as a benchmark for with which to compare other methods.

As one can see in Fig.~\ref{fig:size}, different configurations exhibit different convergence behavior. Axial configurations, $hh$ and $kk$, are more sensitive to the supercell size in the basal plane than the low symmetry basal configurations, $hk$ and $kh$. As divacancy defect states have their largest expansion in a plane perpendicularly to the symmetry of the axis of the defect (cf. Fig.~\ref{fig:divac}) the observed behavior can be explained by the overlap of the wavefunction of the defect and its periodically repeated images, due to the periodic boundary condition. The $hh$ and $kk$ defects extend most in the basal plane, thus, their self-overlap error sensitively depends on the basal plane lateral size of the supercell. For $kh$ and $hk$ the overlap is smaller, since these configurations have their largest expansion in a plane with an angle to the basal plane, thus they are less sensitive to the supercell size in the basal plane. It is clear that the ZPL energies only converges at very large supercell sizes.

Note that, the observed finite size dependence is a combination of several effects, including the convergence of the charge density of 4H-SiC, exponential decrease of the self-interaction wavefunction of the defect, and the relaxation of the strain cased by the defect as the supercell size is increased. In the following subsections, we investigate these effects separately.

\subsubsection{Brillouin zone sampling}
\begin{figure}[h!]
	\includegraphics[width=\columnwidth]{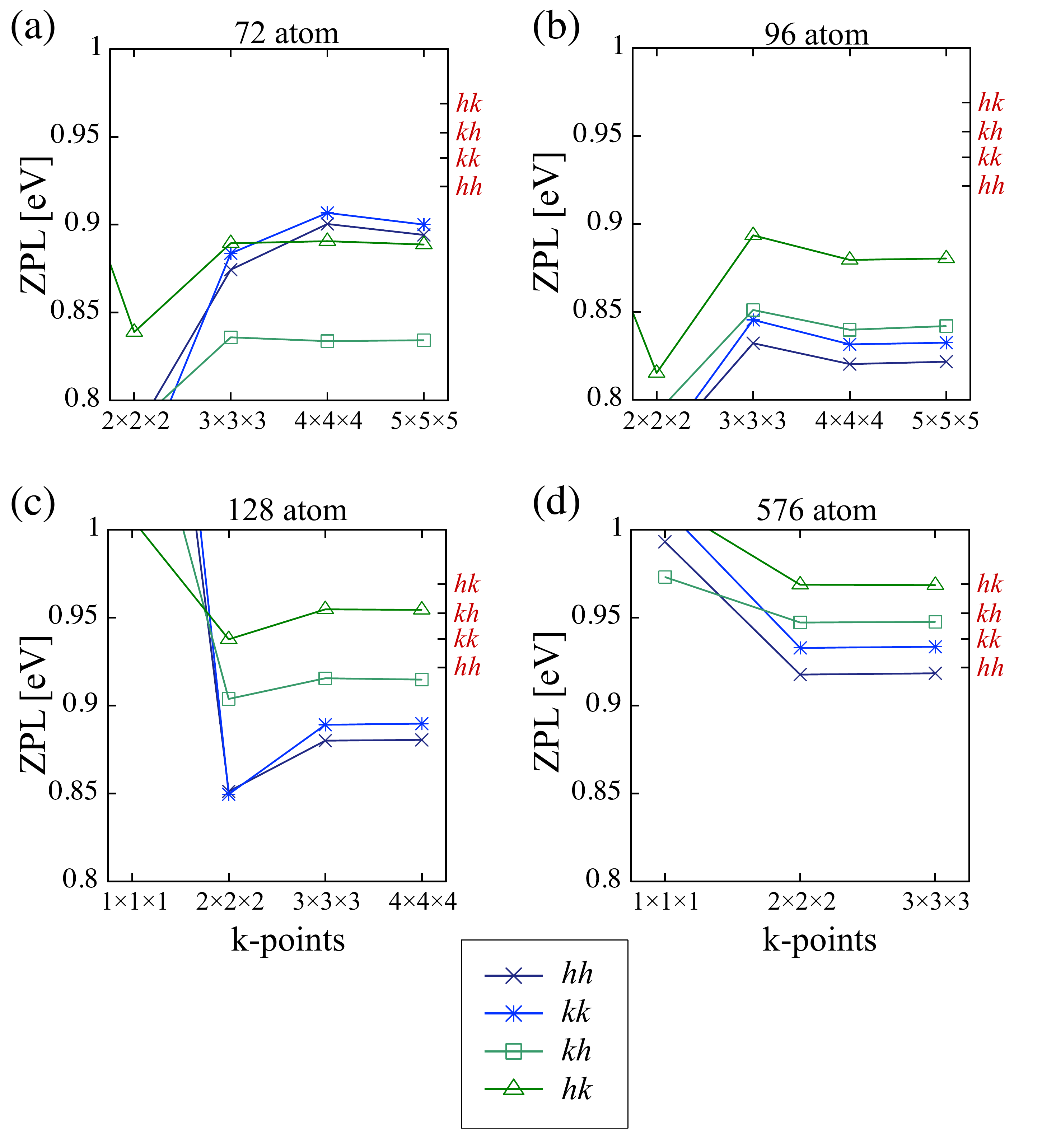}
	\caption{Brillouin zone sampling convergence of ZPL energies for the divacancy defects in different supercell models of 4H-SiC. The right axis shows the converged ZPL for the 2400 supercell.} 
	\label{fig:k-points} 
\end{figure}

In contrast to the straightforward study above of the convergence of ZPL on supercell size in a $\Gamma$-only k-point calculation, we now turn to convergence in smaller cells with higher Brillouin zone sampling. While a smaller supercell will increase errors due to vacancy-vacancy interaction, the aim is to investigate if one can still reach results that are accurate enough with significantly less computational overhead. Fig.~\ref{fig:k-points} shows the PBE ZPL energies for different supercells of hexagonal and rectangular symmetry. As can be seen, the level of BZ sampling is important for all the considered supercells. The order of the ZPL energies of the non-equivalent divacancy configurations largely depends on the k-point convergence. The 576 atom supercell requires a $2 \times 2 \times 2$ BZ sampling to provide the convergent order of PL lines, even though the absolute values are slightly smaller than for those convergent 2400 atom supercell.

\begin{figure}[h!]
    \includegraphics[width=0.8\columnwidth]{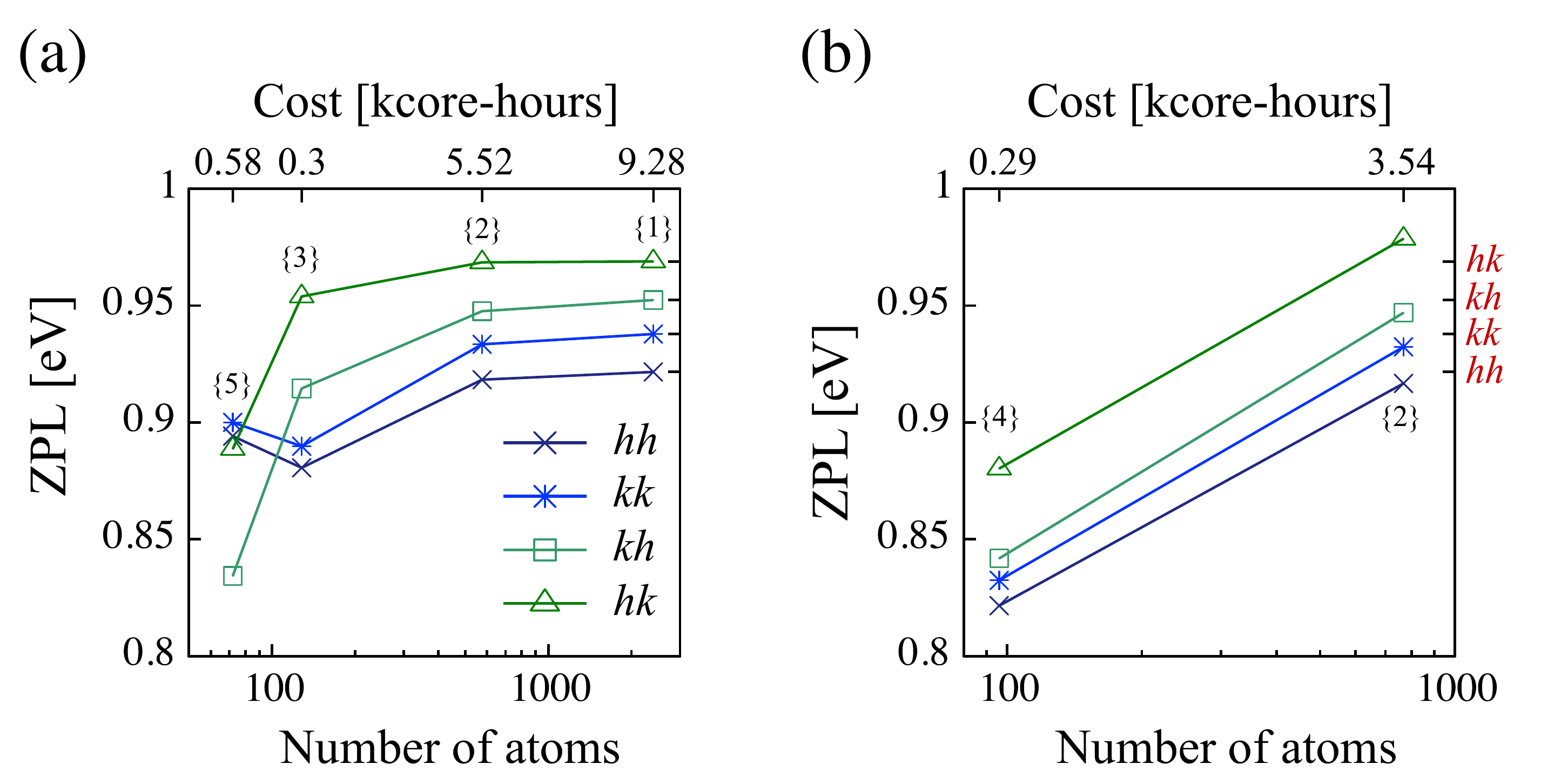}
	\caption{K-point convergent divacancy ZPL energies for different (a) hexagonal and (b) rectangular, as close to cubic as possible, supercells. The right axis shows the converged ZPL for the 2400 supercell. Computational cost is shown on the top axis. \{n\} means $n \times n \times n$ k-points. The presented core-hours are the total amount of computer time needed to calculate one ZPL value on a cluster with 2.2 GHz processors.}
	\label{fig:total} 
\end{figure}

Fig.~\ref{fig:total} shows the final result after k-point convergence of different hexagonal and symmetric supercells. As can be seen, calculations with converged BZ sampling provide the same order for the ZPL line energies as the fully converged 2400 atom supercell for supercells of 96 atoms and larger. On the other hand, the absolute value of the ZPL energies can vary 50 - 100~meV with supercell size. 

These results indicate that careful convergence in k-point density can produce the same order of ZPL energies for smaller supercells than calculations carried out with $\Gamma$-point sampling. The only exception is the hexagonal 72 atom supercell, where either the overlap of the defect states or the point defect caused stress turned to be ruinous.

The presented calculations use a consistent k-point set between the relaxation and ZPL calculation. However, relaxations can be done using only the $\Gamma$-point, with a relative change of the ZPL of only 2\%-4\%. In contrast, the higher k-point sampling is critical for the ZPL energy calculations. This may seem surprising at first, but we speculate that the widening of the bands beyond the zero dispersion of the gamma calculation is important to reproduce the correct physics.

\begin{figure}[h!]
\includegraphics[width=0.7\columnwidth]{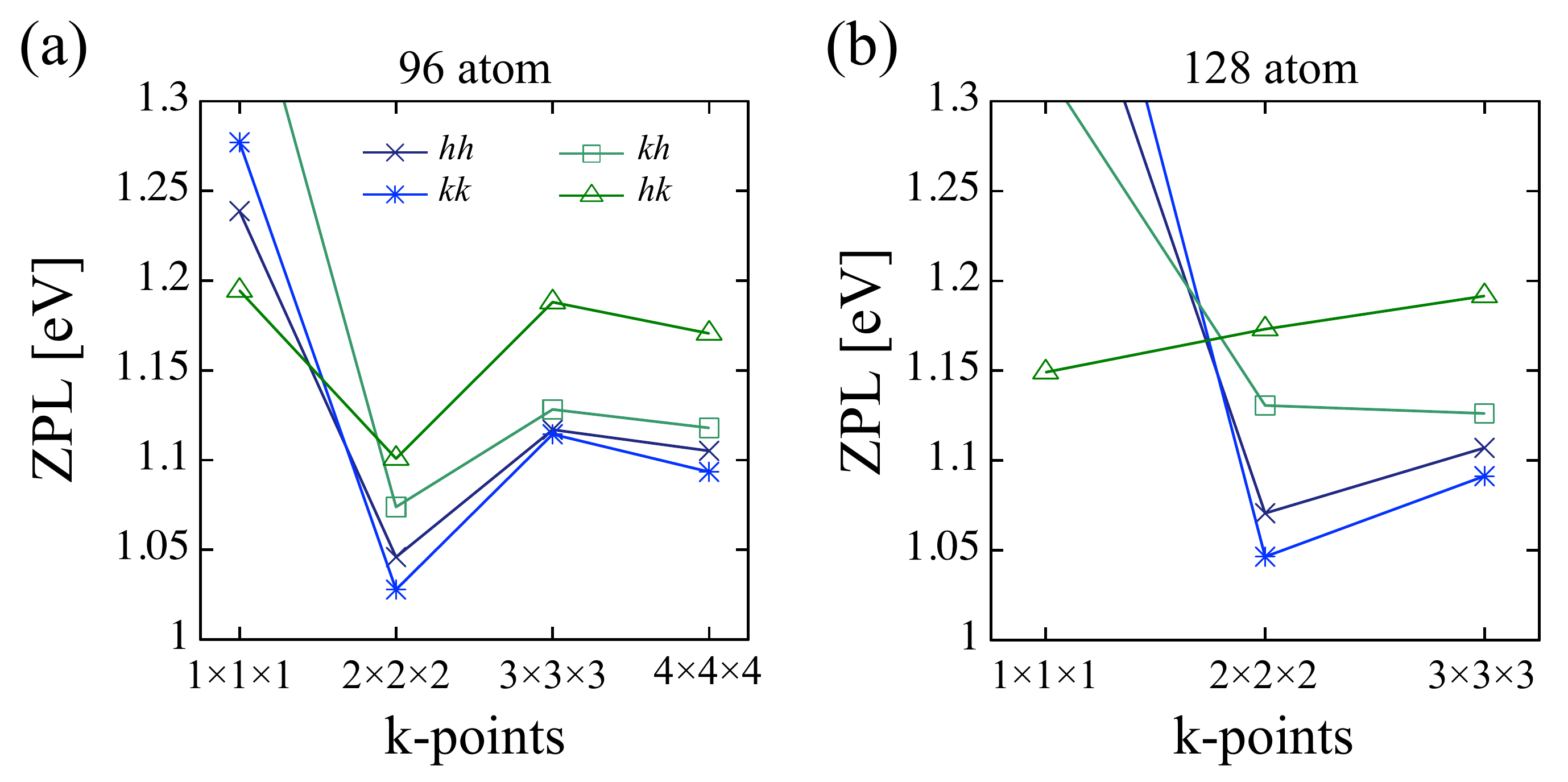}
\caption{Brillouin zone sampling convergence of the HSE06 ZPL energies of divacancy configurations in a rectangular 96 atoms and hexagonal 128 atoms supercell of 4H-SiC. } 
\label{fig:hse} 
\end{figure}

Next, we study the k-point convergence of the ZPL energies calculated by the HSE hybrid functional. Due to the large computational cost of hybrid calculations, we are unable to carry out a study of convergence as thorough as for PBE. Here, we investigate the BZ sampling convergence for the smallest supercell that is sufficient for reproducing the lines in right order. The results for cubic 96 atoms and hexagonal 128 atoms are presented in Fig.~\ref{fig:hse}. As can be seen, the HSE06 functional exhibits similar k-point convergence as the PBE functional. The absolute values of the ZPL are about 0.2 eV larger for the computational heavy HSE06 functional than compared with the PBE functional, also the $hh$ and $kk$ switch order compared to the PBE results. The difference between the ZPL results for the 96 and 128 atoms supercell is smaller for the HSE06 functional than for the PBE functional. This would suggest that the HSE06 functional converges faster with respect to supercell size than the PBE functional.

\subsubsection{Functional and geometry dependence}
\begin{figure}[h!]
\includegraphics[width=0.7\columnwidth]{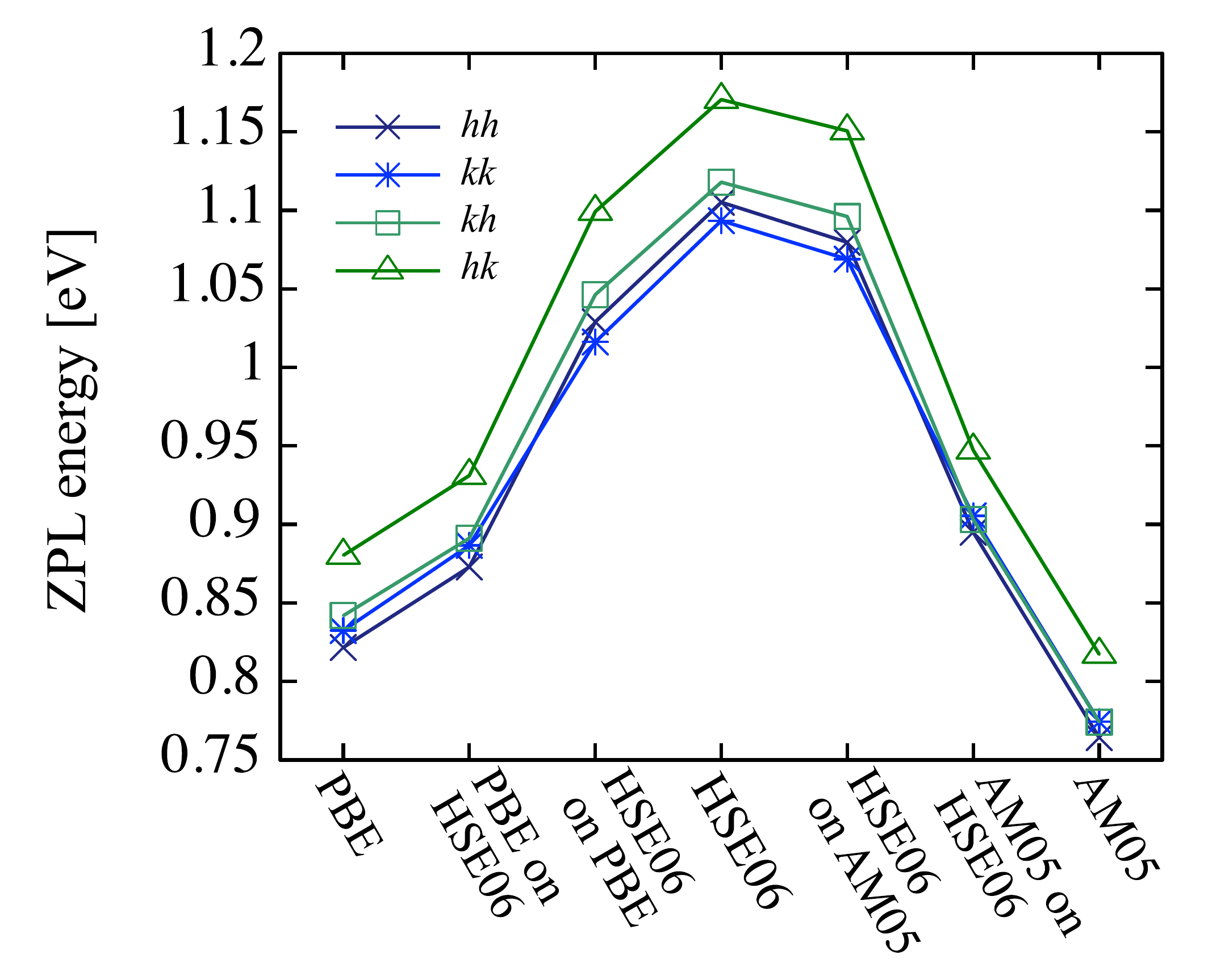}
\caption{Divacancy ZPL energies calculated with various functionals using different strategies. Values for PBE, AM05, and HSE06 have used those functionals for all steps in the calculation. Values specified as 'A on B' utilize the B functional for all relaxations, and the A functional for the final static calculation to determine the ZPL.}
\label{fig:func} 
\end{figure}

Next, we investigate how the choice of the functional and the way how the geometry optimization is performed affect the calculated ZPL energies. Fig.~\ref{fig:func} shows the results of PBE, AM05, and HSE06 ZPL calculations, as well as PBE, AM05, and HSE06 ZPL energies as obtained on different geometries, i.e.\ PBE, AM05, or HSE06 relaxed structures. From these results, it is clear that the functional used for the relaxations has a smaller effect on the results than the functional used in the final static calculation of the ZPL energies. Most importantly, the functional used for relaxation does not change the order of the ZPL lines. While the functional, which is used to calculate the ZPL energies, has an important effect both on the absolute values as well as on the relative positions of the ZPL lines. Concerning the semi-local functionals, use of the PBE functional in all steps of the calculation brings the results closer to the HSE06 values compared to AM05. However, this appears to be related to a cancellation of error effect, since, as shown in Table~\ref{tab:lattice}, the AM05 lattice constants are closer to the experimental ones than PBE.

In addition to the functionals shown in Fig.~\ref{fig:func}, we have also investigated the LDA functional and only comment on the results briefly. Somewhat surprisingly, the LDA results are in significant disagreement with the other functionals, with the final ZPL values of 1.50 for $hh$, 1.55 for $hk$, 1.50 for $kh$, and 0.87 for $kk$. Hence, LDA does not reproduce the experimental order, and, places one of the lines far from the others (0.7 eV away). This can be explained by the underestimated lattice parameter of LDA. Redoing the LDA calculation with the only difference of constraining the lattice constant during relaxation to the PBE value, gives results much closer in agreement with the others functionals (0.77, 0.83, 0.79, 0.78 for $hh$, $hk$, $kh$, and $kk$ respectively). Hence, it appears an incorrect geometry specifically towards a too small lattice constant is disastrous for the end results. This warns against indiscriminate use of LDA in the determination of ZPL lines.
\begin{table}[h!]
\caption{Lattice parameter for 4H-SiC for different functionals compared to experiment.}
\begin{ruledtabular}
\begin{tabular} {c|cc|c|cc}
Lattice parameter & LDA & HSE06 & Exp & AM05 & PBE \\ \hline
a      &   3.059 & 3.071 &  3.073 &  3.077 & 3.094 \\ 
c    &   10.015 & 10.052 &  10.053 & 10.070 &  10.125  \\
\end{tabular}
\end{ruledtabular}
\label{tab:lattice}
\end{table}

\subsubsection{Summary on ZPL line calculations}

\begin{figure}[h!]
\includegraphics[width=0.9\columnwidth]{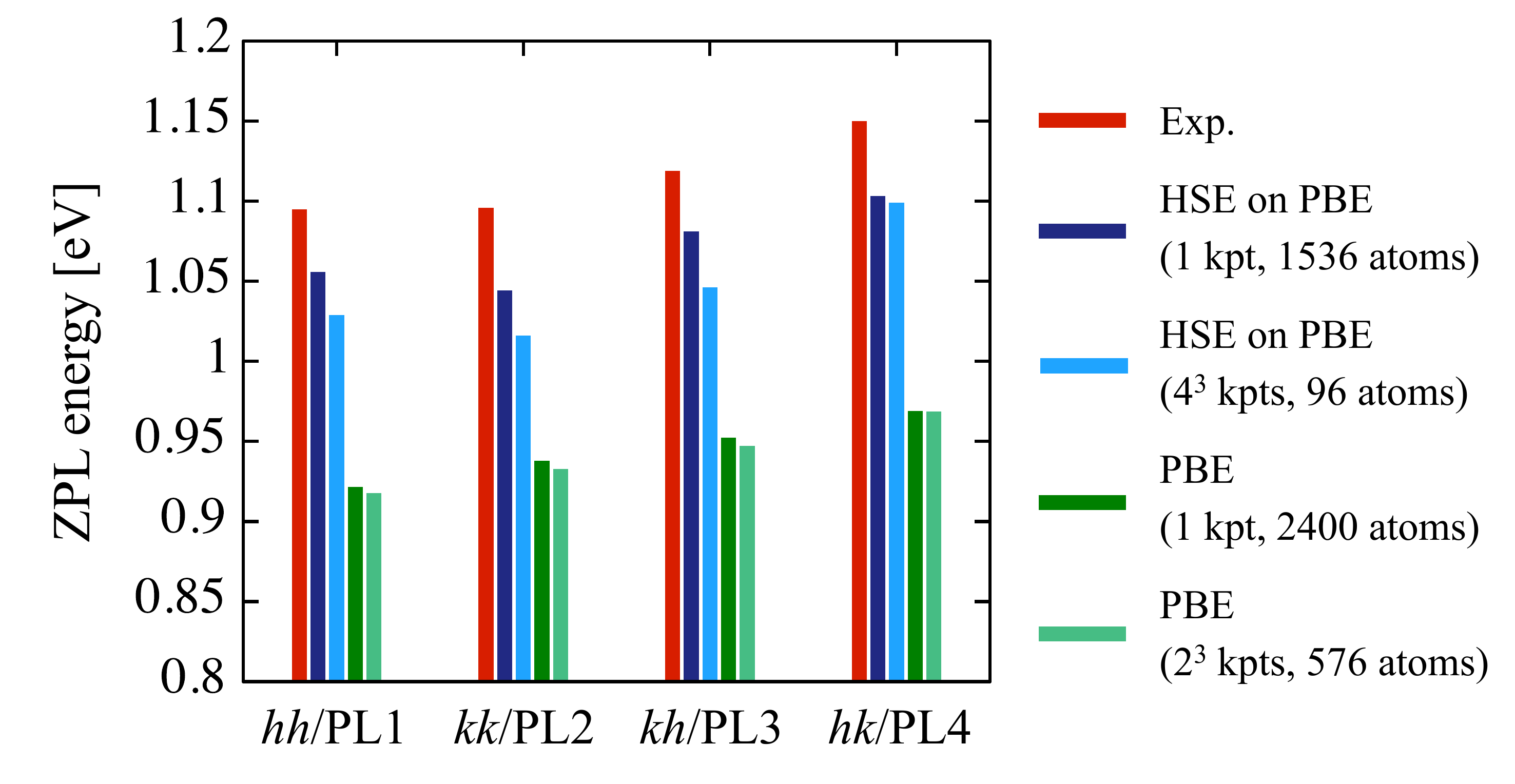}
\caption{Comparison of convergent theoretical ZPL energies with the experimental values\cite{Falk2013}. Identification of the divacancy related PL centers in 4H-SiC is done based on the results of section~\ref{sec:ID}. } 
\label{fig:comp} 
\end{figure}

To summarize our results on ZPL calculations, we depict the experimental and theoretical results of different computational strategies in Fig.~\ref{fig:comp}.  Due to our findings in the previous section, relaxing the structure with PBE, or perhaps preferably AM05, combined with a single self-consistent HSE06 calculation is a fast alternative to running full relaxation with HSE06. As can be seen, in absolute terms the HSE06 functional provides excellent agreement with the experiment values even when PBE optimized geometries are used. PBE ZPL energies exhibit a notable finite-size effect, while the convergent values still fall 20\% below the experimental and HSE06 values. Concerning the order of the PL lines, PBE and HSE06 functionals disagree in the order of the axial divacancy configurations in all the different computational approaches. On the other hand, the experimental energy difference between the axial, high symmetry configurations is only 1~meV, which cannot be achieved by any of ZPL energy calculations strategies, partially due to the neglect of the zero-point energy contribution. Therefore, we cannot decide if the HSE06 or the PBE functionals perform better in non-equivalent configuration identification based on  ZPL energies. On the other hand, due to the non-local nature of HSE06 functional, it may provide a better description of the decaying region of the defect states with the rest of the defect states that can positively affect the predicted order of the ZPL energies. 

The decreased computational time from running PBE functional on 96 atoms (5500 core-hours) instead of 2400 atoms (9300 core-hours) is 40\%. Running HSE on PBE for 1 k-point and 1536 atoms took approximately 36000 core-hours compared with HSE on PBE for $4 \times 4 \times 4$ k-points and 96 atoms which took approximately 27000 core-hours. This is a speedup of 25\%. These core-hours are the total amount of computer time needed to calculate one ZPL value on a cluster with 2.2 GHz processors.

Finally, our results indicate that the computational cost of large HSE06 point defect calculations can be reduced substantially by using PBE or AM05 relaxed geometries with reasonable additional uncertainties both in the absolute values and relative differences of the ZPL energies.

\subsection{Hyperfine field calculations}
\label{sec:hyp}
In this section, we discuss the methodological requirements needed for accurate hyperfine tensor calculations. As it was established previously, the most accurate values can be obtained by HSE06 functional including core state polarization effects\cite{Szasz2013}. Previously, however, no supercell size and BZ sampling tests were carried out.

\begin{figure}[h!]
\includegraphics[width=0.8\columnwidth]{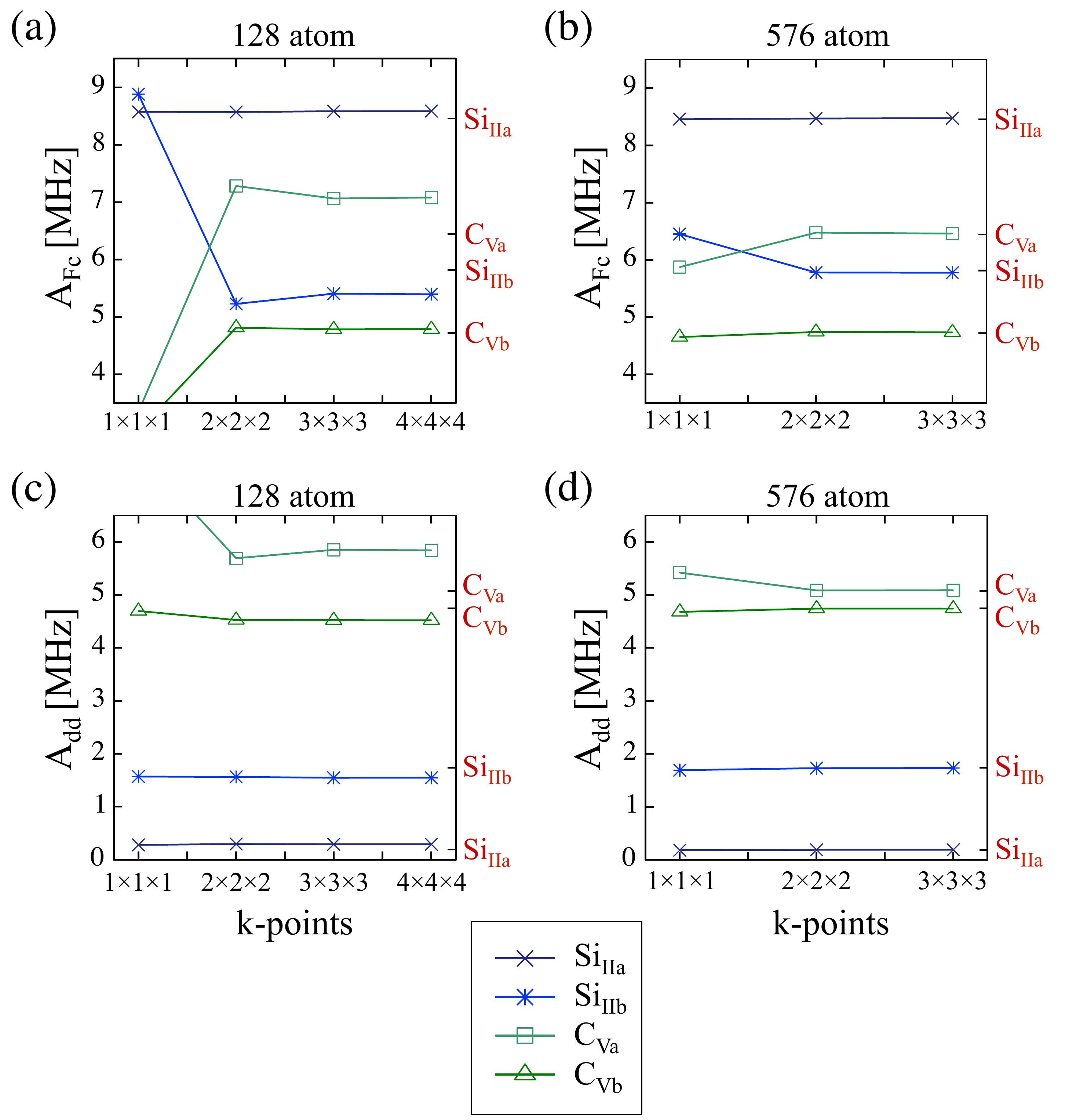}
\caption{Brillouin zone sampling grid size convergence of hyperfine parameters for different $^{29}$Si and $^{13}$C sites in 128 and 576 atom supercells. (a) and (b) show the Fermi contact parameter, while (c) and (d) show the dipole-dipole splitting for 128 and 576 atom supercells, respectively. The considered nuclei sites are marked in Fig.~\ref{fig:divac}(d). The right axis shows the hyperfine parameters for the 2400 supercell. } 
\label{fig:hyp-k} 
\end{figure}

First, we investigate the k-point convergence of hyperfine tensor elements. We use the PBE functional and study different supercell sizes. In the tests we consider two hyperfine parameters, the isotropic Fermi-contact contribution, $A_{\text{Fc}} = \left( A_{xx}+A_{yy}+A_{zz} \right)/3$, and the dipole-dipole splitting, $A_{\text{dd}} = \left| \left(A_{xx} + A_{yy} \right) / 2 - A_{zz} \right|$.  The obtained convergence curves for a set of $^{29}$Si and $^{13}$C sites close to a $hh$ divacancy are depicted in Fig.~\ref{fig:hyp-k}. As one can see different sites exhibit different convergence behavior, which can be explained by the fact that the defect has different extension in different directions. In the basal plane, the extension is larger thus the defect states may require denser k-point grid in this direction. Furthermore, one can see that the criteria of complete convergence is similar to the criteria obtained for the ZPL energies.  

\begin{figure}[h!]
\includegraphics[width=0.4\columnwidth]{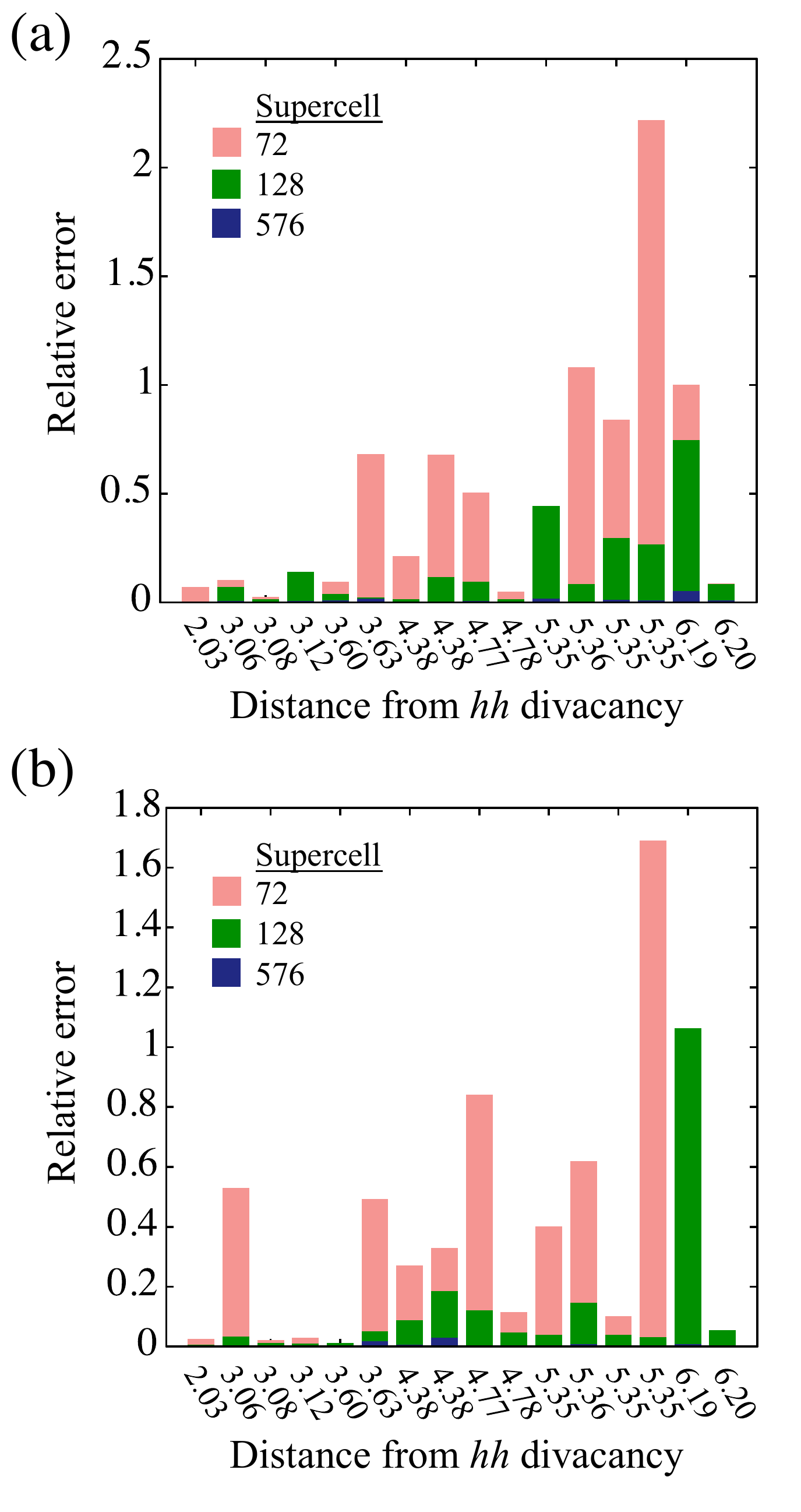}
\caption{Relative error of the calculated (a) Fermi contact and (b) dipole-dipole hyperfine interactions strength for various sites around a $hh$ divacancy in 4H-SiC. On the horizontal axis, the distances of the considered sites from the silicon vacancy site of the divacancy  are given. Calculations were carried out in 72, 128, and 576 atom supercell by using the PBE functional.} 
\label{fig:hyp} 
\end{figure}

Next, we investigate the supercell size dependence of $A_{\text{Fc}}$ and $A_{\text{dd}}$ hyperfine parameters. To do so we calculate the deviation of hyperfine parameters for numerous sites from the values obtained in a 2400 atom supercell calculation. In Fig.~\ref{fig:hyp}, relative errors of $A_{\text{Fc}}$ and $A_{\text{dd}}$ are shown for three smaller supercells. For all considered sites, the distance from the silicon vacancy site of the divacancy is also provided in the figure. As one can see, the relative error in the Fermi contact term increases dramatically with increasing distance of the divacancy and nuclear spins. Similar tendency can be observed for the dipole - dipole hyperfine term. Furthermore, relative errors drastically decrease with increasing supercell size. The mean relative errors for different supercells are provided in Table~\ref{tab:hypMAE}. Note that substantial errors were obtained even in the 128 atom supercell, while the 576 atom supercell results are nearly identical with the absolutely convergent 2400 atom supercell results. These observations can be explained by the overlap of the state from the defect and its replicas. As further hyperfine interaction of the nuclei sensitively depends on the spin density localization in the farther neighbor shells, where defect state overlap can occur, they exhibit an enhanced finite size effect. 

\begin{table}[h!]
\caption{Mean relative error (MRE) of the hyperfine interaction parameters presented in Fig.~\ref{fig:hyp}. Deviation measured from the hyperfine parameters obtained in absolutely convergent 2400 atom supercell calculations.}
\begin{ruledtabular}
\begin{tabular} {c|cc}
Supercell &   MRE of $A_{\text{Fc}}$ [\%] &  MRE of $A_{\text{dd}}$ [\%] \\ \hline
72      &    48.0    &    72.6  \\ 
128    &    15.3    &    25.5  \\
576    &    0.99    &    0.81  \\
\end{tabular}
\end{ruledtabular}
\label{tab:hypMAE}
\end{table}

In summary, our results indicate that accurate hyperfine field calculations require supercells as large as 576 atoms or $\approx 20$~\AA\ lateral size. In smaller supercells, such as 128 atom supercell, only closest hyperfine field of the nuclei can be determined with reasonable accuracy. 

\subsection{Defect induced stress in supercell calculations}

Point defects induce a distortion in their host crystal, which relaxes with increasing distances from the point defects. In finite supercell models, the size of the model is usually not sufficient to accommodate completely the induced strain field around a point defect, thus the atomic configuration of the defect cannot reach the single defect configuration in such calculations. This finite size effect also manifest itself as an artificial stress appears at the supercell surfaces. Here, we investigate how the induced stress relaxes with increasing supercell size.

\begin{figure}[h!]
\includegraphics[width=0.7\columnwidth]{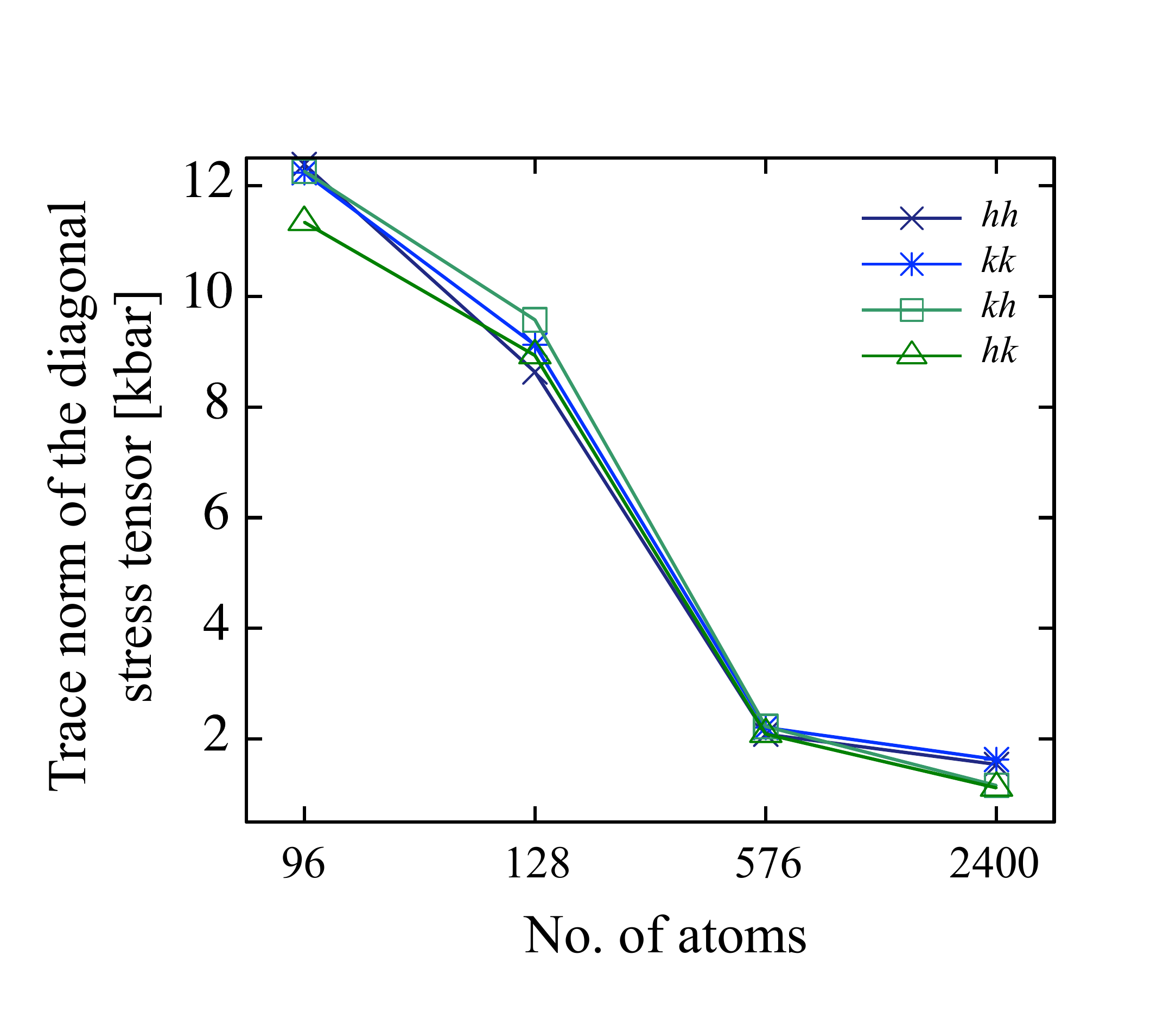}
\caption{Trace norm of the diagonalized stress tensor measured at the borders of different supercells of 4H-SiC embedding single divacancy defects.} 
\label{fig:stress} 
\end{figure}

As can be seen in Fig.~\ref{fig:stress}, the stress indeed relaxes with increasing supercell size, however, it does not reach zero even in the largest 2400 atom supercell. Furthermore, the small difference between the 576 and 2400 atom supercell results may suggest that 576 atom supercell is convergent in terms of stress. 

In order to make a rough estimate how the observed stress affects the ZPL energies, we imagine that the NV center is subject to 10~kbar external pressure, which is approximately the difference of the stress observed in the smallest and largest supercells. By using the experimental pressure dependence of the ZPL energy, 0.575 meV/kbar \cite{Doherty2014}, we obtain 5.75~meV. In SiC the effect could be larger, due to the smaller bulk modulus of SiC, however, presumably still remains in the order of 10~meV. Therefore, we conclude that the stress has a minor effect on the calculated ZPL energies. 

\section{Identification of divacancy related zero-phonon lines in 4H-SiC}
\label{sec:ID}

In this section, we present an example of how accurate ZPL, ZFS, and hyperfine field splitting calculations can be used for identification of the divacancy configurations in 4H-SiC. We now consider a hypothetical case where one has been given the experimental ZPL lines in Table III for the four different defect configurations. Our aim is to identify the type of defect (vacancy, divacancy, interstitial, etc.), and match each line to a corresponding configuration ($hh$, $hk$, $kh$, and $kk$.) 

First, the variation of ZPL energies between different defect types are usually on a scale $>$ 100 meV. Hence, it should be straightforward to make the identification of defect type with access to k-point-converged results for HSE06 for the 96 atom unit cell with the geometry converged using PBE or AM05. Even the PBE results for 96 atom supercell could work to identify the defect type, even though the absolute error is larger, the relative error is on the same scale (cf. Fig.~\ref{fig:k-points}(b) and Fig.~\ref{fig:hse}(a)). Next, we turn to the identification of the defect configuration corresponding to each line. The conclusion from the present work is that the accuracy of the ZPL energies with the methods described here are not of sufficient quality to identify the configurations from them alone (cf. Fig.~\ref{fig:comp}). As has been discussed above, while PBE predicts the correct order for the ZPL, HSE06 does not. However, if we in addition to the ZPL also have experimental results for the ZFS and the hyperfine tensor, they can be used to aid the identification. These quantities require calculations using HSE06 with one k-point and a supercell converged in size (as described in Sec.~\ref{sec:hyp} and Ref.~\cite{Szasz2013,Ivady2014}). For the divacancy, we use 1538 atom supercell ($8 \times 8 \times 3$) with $24.7 \times 24.7 \times 30.4$~\AA$^{3}$ volume, with $\Gamma$-point sampling. This calculation is converged in the $c$-axis direction and nearly converged in basal directions. To reduce the demand of such calculations, we use PBE relaxed atomic configurations and single self-consistent HSE06 calculations. This calculation provides accurate hyperfine and ZFS results and also increases the accuracy of the ZPL lines. The results in Table~\ref{tab:ID} show that the added data and accuracy is sufficient to identify each line. The ZFS result for $kh$ is far from experiment, this could be due to the neglect of the spin-orbit contribution.

Hence, we suggest a two-step process for defect identification. First, a database is needed for ZPL lines for all relevant defects and configurations, produced by PBE calculations for 96 atom supercells with $4 \times 4 \times 4$ BZ-sampling at a cost of 300 core hours/defect configuration on a cluster with 2.2 GHz processors. This database hopefully allows the identification of different defect type. Second, to identify the defect configuration requires additional calculations of ZFS and the hyperfine parameters. To produce accurate results, an HSE06 calculation on 1536 atom supercell, with $\Gamma$-point sampling, relaxed using PBE functional is needed. This also produces more accurate ZPL at a total cost of ca 35000 core hours/defect configuration. 

To identify the defect, first, we chose an affordable but sufficiently accurate method to calculate all the necessary parameters. By studying the ZPL data presented in Fig.~\ref{fig:comp}, one can identify $kh$ and $hk$ either from the PBE or HSE06 functional results. Even the results from a small supercell with high k-point set can identify the different configuration. But the $hh$ and $kk$ identification is conflicting between the functionals, then one needs other properties such as ZFS or hyperfine tensor to be sure. As we have seen, the HSE06 functional is required for accurate hyperfine tensor calculations\cite{Szasz2013} and also provides better results for the ZPL energies, due to its non-local nature. Furthermore, large supercells have many advantages that affect all the considered quantities and are needed for calculating accurate hyperfine tensor, even doe they are more computational demanding. Here, we use 1538 atom supercell ($8 \times 8 \times 3$) with $24.7 \times 24.7 \times 30.4$~\AA$^{3}$ volume, with $\Gamma$-point sampling. This calculation is convergent in the $c$-axis direction and nearly convergent in basal directions. In this non-completely convergent calculations small uncertainties are expected, e.g. 10~meV in the ZPL energies, see Fig.~\ref{fig:size}, especially when axial and basal configurations are compared. To reduce the demand of such calculations, we use PBE relaxed atomic configurations and single self-consistent HSE06 calculations. ZFS is calculated using the method presented in Ref.~\cite{Ivady2014}. This property is assumed to have the same convergence as hyperfine field. The results of these calculations are summarized in Table~\ref{tab:ID}. 

\begin{table}[h!]
\caption{Identification of divacancy related PL and ESR centers in 4H-SiC.}
\begin{ruledtabular}
\begin{tabular} {c|cc||cc|cc|cc}
Configuration &   PL line\cite{Koehl11,Falk2013} &  ESR\cite{Baranov2005,Son2006}    & Calc.  ZPL  & Exp. ZPL\cite{Falk2013}  & Calc. ZFS & Exp. ZFS\cite{Falk2013} & Calc. $A_z$ & Exp. $A_z$\cite{FalkPRL2015} \\ \hline
$hh$      &    PL1   &    P6b & 1.056  & 1.095  & 1.329  &  1.336 & 9.06  & 9.2  \\ 
$kk$    &    PL2    &    P6'b  &  1.044 &  1.096 & 1.307  &  1.305 & 9.99  & 10.0 \\
$kh$    &    PL3    &   P7'b  & 1.081 &  1.119   & 1.314  &   1.222 & -- &  --  \\
$hk$    &    PL4    &   P7b   & 1.103 &   1.150  & 1.363  &  1.334  & -- &  --  \\
\end{tabular}
\end{ruledtabular}
\label{tab:ID}
\end{table}

Comparing with prior work, our identification agrees with that by Falk \emph{et al.}\cite{Falk2014}, which uses the ZFS parameter and \emph{ab initio} simulations. The ZFS parameter was calculated using the data from a 1200 atom supercell, with $\Gamma$-point sampling, using the PBE functional. Our computational results also agree with a calculation by Gordon \emph{et al.}\cite{Gordon2015} using the HSE06 functional on a 96 atom supercell with $2 \times 2 \times 2$ k-point grid. As the authors remarked it is not possible to use only ZPL to identify the different configurations due to the low accuracy of the calculations. The HSE06 functional predicts the wrong order and as discussed in the present work more accurate calculations do not resolve this but additional properties is needed for a correct identification. 

\section{Summary}
This work discusses how to appropriately use \emph{ab initio} calculations to facilitate identification of defect types and configurationsin semiconductors. Specifically, we have shown how to correctly identify the different non-equivalent divacancy configurations in 4H-SiC using ZPL, ZFS, and hyperfine field splitting calculations.

The value and order of the zero-phonon lines are dependent on the choice of functional, the way the geometry is optimized, supercell size, and k-points density. A comparably small supercell that has been converged with respect to the number of k-points produces sufficient accurate results at lower computational cost than the typical setup of a large supercell with one k-point. The absolute value of the zero-phonon line energy depends strongly on the lattice constant. A smaller lattice constant gives larger zero-phonon line energy and vice versa. The functional affects both the order and absolute value. It turns out that using only ZPL data is not enough to successfully identify the different non-equivalent configurations, but ZFS and hyperfine field can provide additional information.

For hyperfine field, the size of the supercell is the most important factor. Close to the defect, the values do not vary as the supercell size changes. But as the supercell size decreases, the hyperfine field gets less accurate the further away from the defect one calculates it.

The approach that produces both most accurate zero-phonon lines and hyperfine field values, at an affordable cost, is to use a large enough supercell that only $\Gamma$-point sampling is needed, relax it with PBE or AM05 functional, and run a single self-consistent HSE06 calculation. Using this proposed algorithm, we have shown how to correctly identify the different non-equivalent divacancy configurations in 4H-SiC.

\section*{Acknowledgments} 
Support from the Swedish Government Strategic Research Areas in Materials Science on Functional Materials at Link\"oping University (Faculty Grant SFO-Mat-LiU No. 2009-00971) and the Swedish e-Science Centre (SeRC), Knut \& Alice Wallenberg Foundation New States of Matter 2014-2019 (COTXS), the Swedish Research Council (VR) Grants No.\ 621-2011-4426 and  2016-04810, the Swedish National Infrastructure for Computing Grants No. SNIC 001/12-275 and No. SNIC 2013/1-331, and the ``Lend\"ulet program" of Hungarian Academy of Sciences is acknowledged. Support  provided by the Ministry of Education and Science of the Russian Federation (Grant No. 14.Y26.31.0005) is gratefully acknowledged. Use of the Center for Nanoscale Materials was supported by the U. S. Department of Energy, Office of Science, Office of Basic Energy Sciences, under Contract No. DE-AC02-06CH11357.

\bibliographystyle{apsrev4-1}
\bibliography{references}

\begin{thebibliography}{63}%
\makeatletter
\providecommand \@ifxundefined [1]{%
 \@ifx{#1\undefined}
}%
\providecommand \@ifnum [1]{%
 \ifnum #1\expandafter \@firstoftwo
 \else \expandafter \@secondoftwo
 \fi
}%
\providecommand \@ifx [1]{%
 \ifx #1\expandafter \@firstoftwo
 \else \expandafter \@secondoftwo
 \fi
}%
\providecommand \natexlab [1]{#1}%
\providecommand \enquote  [1]{``#1''}%
\providecommand \bibnamefont  [1]{#1}%
\providecommand \bibfnamefont [1]{#1}%
\providecommand \citenamefont [1]{#1}%
\providecommand \href@noop [0]{\@secondoftwo}%
\providecommand \href [0]{\begingroup \@sanitize@url \@href}%
\providecommand \@href[1]{\@@startlink{#1}\@@href}%
\providecommand \@@href[1]{\endgroup#1\@@endlink}%
\providecommand \@sanitize@url [0]{\catcode `\\12\catcode `\$12\catcode
  `\&12\catcode `\#12\catcode `\^12\catcode `\_12\catcode `\%12\relax}%
\providecommand \@@startlink[1]{}%
\providecommand \@@endlink[0]{}%
\providecommand \url  [0]{\begingroup\@sanitize@url \@url }%
\providecommand \@url [1]{\endgroup\@href {#1}{\urlprefix }}%
\providecommand \urlprefix  [0]{URL }%
\providecommand \Eprint [0]{\href }%
\providecommand \doibase [0]{http://dx.doi.org/}%
\providecommand \selectlanguage [0]{\@gobble}%
\providecommand \bibinfo  [0]{\@secondoftwo}%
\providecommand \bibfield  [0]{\@secondoftwo}%
\providecommand \translation [1]{[#1]}%
\providecommand \BibitemOpen [0]{}%
\providecommand \bibitemStop [0]{}%
\providecommand \bibitemNoStop [0]{.\EOS\space}%
\providecommand \EOS [0]{\spacefactor3000\relax}%
\providecommand \BibitemShut  [1]{\csname bibitem#1\endcsname}%
\let\auto@bib@innerbib\@empty
\bibitem [{\citenamefont {Lannoo}(2012)}]{BookPointDefect1}%
  \BibitemOpen
  \bibfield  {author} {\bibinfo {author} {\bibfnamefont {M.}~\bibnamefont
  {Lannoo}},\ }\href@noop {} {\emph {\bibinfo {title} {Point defects in
  semiconductors I: theoretical aspects}}},\ Vol.~\bibinfo {volume} {22}\
  (\bibinfo  {publisher} {Springer Science \& Business Media},\ \bibinfo {year}
  {2012})\BibitemShut {NoStop}%
\bibitem [{\citenamefont {Bourgoin}(2012)}]{BookPointDefect2}%
  \BibitemOpen
  \bibfield  {author} {\bibinfo {author} {\bibfnamefont {J.}~\bibnamefont
  {Bourgoin}},\ }\href@noop {} {\emph {\bibinfo {title} {Point defects in
  Semiconductors II: Experimental aspects}}},\ Vol.~\bibinfo {volume} {35}\
  (\bibinfo  {publisher} {Springer Science \& Business Media},\ \bibinfo {year}
  {2012})\BibitemShut {NoStop}%
\bibitem [{\citenamefont {Childress}\ \emph {et~al.}(2006)\citenamefont
  {Childress}, \citenamefont {Taylor}, \citenamefont {S\o{}rensen},\ and\
  \citenamefont {Lukin}}]{Childress:PRL2006}%
  \BibitemOpen
  \bibfield  {author} {\bibinfo {author} {\bibfnamefont {L.}~\bibnamefont
  {Childress}}, \bibinfo {author} {\bibfnamefont {J.~M.}\ \bibnamefont
  {Taylor}}, \bibinfo {author} {\bibfnamefont {A.~S.}\ \bibnamefont
  {S\o{}rensen}}, \ and\ \bibinfo {author} {\bibfnamefont {M.~D.}\ \bibnamefont
  {Lukin}},\ }\href {\doibase 10.1103/PhysRevLett.96.070504} {\bibfield
  {journal} {\bibinfo  {journal} {Phys. Rev. Lett.}\ }\textbf {\bibinfo
  {volume} {96}},\ \bibinfo {pages} {070504} (\bibinfo {year}
  {2006})}\BibitemShut {NoStop}%
\bibitem [{\citenamefont {Aharonovich}\ \emph {et~al.}(2009)\citenamefont
  {Aharonovich}, \citenamefont {Castelletto}, \citenamefont {Simpson},
  \citenamefont {Stacey}, \citenamefont {McCallum}, \citenamefont {Greentree},\
  and\ \citenamefont {S.}}]{Aharonovich:NL2009}%
  \BibitemOpen
  \bibfield  {author} {\bibinfo {author} {\bibfnamefont {I.}~\bibnamefont
  {Aharonovich}}, \bibinfo {author} {\bibfnamefont {S.}~\bibnamefont
  {Castelletto}}, \bibinfo {author} {\bibfnamefont {D.~A.}\ \bibnamefont
  {Simpson}}, \bibinfo {author} {\bibfnamefont {A.}~\bibnamefont {Stacey}},
  \bibinfo {author} {\bibfnamefont {J.}~\bibnamefont {McCallum}}, \bibinfo
  {author} {\bibfnamefont {A.~D.}\ \bibnamefont {Greentree}}, \ and\ \bibinfo
  {author} {\bibfnamefont {P.}~\bibnamefont {S.}},\ }\href@noop {} {\bibfield
  {journal} {\bibinfo  {journal} {Nano Letters}\ }\textbf {\bibinfo {volume}
  {9}},\ \bibinfo {pages} {3191} (\bibinfo {year} {2009})}\BibitemShut
  {NoStop}%
\bibitem [{\citenamefont {Kolesov}\ \emph {et~al.}(2012)\citenamefont
  {Kolesov}, \citenamefont {Xia}, \citenamefont {Reuter}, \citenamefont
  {St\"{o}hr}, \citenamefont {Zappe}, \citenamefont {Meijer}, \citenamefont
  {Hemmer},\ and\ \citenamefont {Wrachtrup}}]{Kolesov2012}%
  \BibitemOpen
  \bibfield  {author} {\bibinfo {author} {\bibfnamefont {R.}~\bibnamefont
  {Kolesov}}, \bibinfo {author} {\bibfnamefont {K.}~\bibnamefont {Xia}},
  \bibinfo {author} {\bibfnamefont {R.}~\bibnamefont {Reuter}}, \bibinfo
  {author} {\bibfnamefont {R.}~\bibnamefont {St\"{o}hr}}, \bibinfo {author}
  {\bibfnamefont {A.}~\bibnamefont {Zappe}}, \bibinfo {author} {\bibfnamefont
  {J.}~\bibnamefont {Meijer}}, \bibinfo {author} {\bibfnamefont {P.~R.}\
  \bibnamefont {Hemmer}}, \ and\ \bibinfo {author} {\bibfnamefont
  {J.}~\bibnamefont {Wrachtrup}},\ }\href {http://dx.doi.org/10.1038/ncomms2034
  10.1038/ncomms2034
  http://www.nature.com/articles/ncomms2034\#supplementary-information}
  {\bibfield  {journal} {\bibinfo  {journal} {Nature Communications}\ }\textbf
  {\bibinfo {volume} {3}},\ \bibinfo {pages} {1029} (\bibinfo {year}
  {2012})}\BibitemShut {NoStop}%
\bibitem [{\citenamefont {Castelletto}\ \emph {et~al.}(2014)\citenamefont
  {Castelletto}, \citenamefont {Johnson}, \citenamefont {Iv\'ady},
  \citenamefont {Stavrias}, \citenamefont {Umeda}, \citenamefont {Gali},\ and\
  \citenamefont {Ohshima}}]{NatMat14}%
  \BibitemOpen
  \bibfield  {author} {\bibinfo {author} {\bibfnamefont {S.}~\bibnamefont
  {Castelletto}}, \bibinfo {author} {\bibfnamefont {B.~C.}\ \bibnamefont
  {Johnson}}, \bibinfo {author} {\bibfnamefont {V.}~\bibnamefont {Iv\'ady}},
  \bibinfo {author} {\bibfnamefont {N.}~\bibnamefont {Stavrias}}, \bibinfo
  {author} {\bibfnamefont {T.}~\bibnamefont {Umeda}}, \bibinfo {author}
  {\bibfnamefont {A.}~\bibnamefont {Gali}}, \ and\ \bibinfo {author}
  {\bibfnamefont {T.}~\bibnamefont {Ohshima}},\ }\href {\doibase
  10.1038/nmat3806} {\bibfield  {journal} {\bibinfo  {journal} {Nat. Mater.}\
  }\textbf {\bibinfo {volume} {13}},\ \bibinfo {pages} {151} (\bibinfo {year}
  {2014})}\BibitemShut {NoStop}%
\bibitem [{Aha(2016)}]{Aharonovich2016}%
  \BibitemOpen
  \href {http://dx.doi.org/10.1038/nphoton.2016.186 10.1038/nphoton.2016.186}
  {\bibfield  {journal} {\bibinfo  {journal} {Nat Photon}\ }\textbf {\bibinfo
  {volume} {10}},\ \bibinfo {pages} {631} (\bibinfo {year} {2016})}\BibitemShut
  {NoStop}%
\bibitem [{\citenamefont {Jelezko}\ and\ \citenamefont
  {Wrachtrup}(2006)}]{Jelezko:PSS2006}%
  \BibitemOpen
  \bibfield  {author} {\bibinfo {author} {\bibfnamefont {F.}~\bibnamefont
  {Jelezko}}\ and\ \bibinfo {author} {\bibfnamefont {J.}~\bibnamefont
  {Wrachtrup}},\ }\href {\doibase 10.1002/pssa.200671403} {\bibfield  {journal}
  {\bibinfo  {journal} {physica status solidi (a)}\ }\textbf {\bibinfo {volume}
  {203}},\ \bibinfo {pages} {3207} (\bibinfo {year} {2006})}\BibitemShut
  {NoStop}%
\bibitem [{\citenamefont {Hanson}\ and\ \citenamefont
  {Awschalom}(2008)}]{Hanson:Nature2008}%
  \BibitemOpen
  \bibfield  {author} {\bibinfo {author} {\bibfnamefont {R.}~\bibnamefont
  {Hanson}}\ and\ \bibinfo {author} {\bibfnamefont {D.~D.}\ \bibnamefont
  {Awschalom}},\ }\href@noop {} {\bibfield  {journal} {\bibinfo  {journal}
  {Nature (London)}\ }\textbf {\bibinfo {volume} {453}},\ \bibinfo {pages}
  {1043} (\bibinfo {year} {2008})}\BibitemShut {NoStop}%
\bibitem [{\citenamefont {Awschalom}\ \emph {et~al.}(2013)\citenamefont
  {Awschalom}, \citenamefont {Bassett}, \citenamefont {Dzurak}, \citenamefont
  {Hu},\ and\ \citenamefont {Petta}}]{Awschalom2013}%
  \BibitemOpen
  \bibfield  {author} {\bibinfo {author} {\bibfnamefont {D.~D.}\ \bibnamefont
  {Awschalom}}, \bibinfo {author} {\bibfnamefont {L.~C.}\ \bibnamefont
  {Bassett}}, \bibinfo {author} {\bibfnamefont {A.~S.}\ \bibnamefont {Dzurak}},
  \bibinfo {author} {\bibfnamefont {E.~L.}\ \bibnamefont {Hu}}, \ and\ \bibinfo
  {author} {\bibfnamefont {J.~R.}\ \bibnamefont {Petta}},\ }\href {\doibase
  10.1126/science.1231364} {\bibfield  {journal} {\bibinfo  {journal}
  {Science}\ }\textbf {\bibinfo {volume} {339}},\ \bibinfo {pages} {1174}
  (\bibinfo {year} {2013})}\BibitemShut {NoStop}%
\bibitem [{\citenamefont {du~Preez}(1965)}]{duPreez:1965}%
  \BibitemOpen
  \bibfield  {author} {\bibinfo {author} {\bibfnamefont {L.}~\bibnamefont
  {du~Preez}},\ }\href@noop {} {Ph.D. thesis},\ \bibinfo  {school} {University
  of Witwatersrand} (\bibinfo {year} {1965})\BibitemShut {NoStop}%
\bibitem [{\citenamefont {Balasubramanian}\ \emph {et~al.}(2009)\citenamefont
  {Balasubramanian}, \citenamefont {Neumann}, \citenamefont {Twitchen},
  \citenamefont {Markham}, \citenamefont {Kolesov}, \citenamefont {Mizuochi},
  \citenamefont {Isoya}, \citenamefont {Achard}, \citenamefont {Beck},
  \citenamefont {Tissler}, \citenamefont {Jacques}, \citenamefont {Hemmer},
  \citenamefont {Jelezko},\ and\ \citenamefont
  {Wrachtrup}}]{Balasubramanian:NatMat2009}%
  \BibitemOpen
  \bibfield  {author} {\bibinfo {author} {\bibfnamefont {G.}~\bibnamefont
  {Balasubramanian}}, \bibinfo {author} {\bibfnamefont {P.}~\bibnamefont
  {Neumann}}, \bibinfo {author} {\bibfnamefont {D.}~\bibnamefont {Twitchen}},
  \bibinfo {author} {\bibfnamefont {M.}~\bibnamefont {Markham}}, \bibinfo
  {author} {\bibfnamefont {R.}~\bibnamefont {Kolesov}}, \bibinfo {author}
  {\bibfnamefont {N.}~\bibnamefont {Mizuochi}}, \bibinfo {author}
  {\bibfnamefont {J.}~\bibnamefont {Isoya}}, \bibinfo {author} {\bibfnamefont
  {J.}~\bibnamefont {Achard}}, \bibinfo {author} {\bibfnamefont
  {J.}~\bibnamefont {Beck}}, \bibinfo {author} {\bibfnamefont {J.}~\bibnamefont
  {Tissler}}, \bibinfo {author} {\bibfnamefont {V.}~\bibnamefont {Jacques}},
  \bibinfo {author} {\bibfnamefont {P.~R.}\ \bibnamefont {Hemmer}}, \bibinfo
  {author} {\bibfnamefont {F.}~\bibnamefont {Jelezko}}, \ and\ \bibinfo
  {author} {\bibfnamefont {J.}~\bibnamefont {Wrachtrup}},\ }\href
  {http://dx.doi.org/10.1038/nmat2420} {\bibfield  {journal} {\bibinfo
  {journal} {Nature Mater.}\ }\textbf {\bibinfo {volume} {8}},\ \bibinfo
  {pages} {383} (\bibinfo {year} {2009})}\BibitemShut {NoStop}%
\bibitem [{\citenamefont {Buckley}\ \emph {et~al.}(2010)\citenamefont
  {Buckley}, \citenamefont {Fuchs}, \citenamefont {Bassett},\ and\
  \citenamefont {Awschalom}}]{Awschalom:Nature2010}%
  \BibitemOpen
  \bibfield  {author} {\bibinfo {author} {\bibfnamefont {B.~B.}\ \bibnamefont
  {Buckley}}, \bibinfo {author} {\bibfnamefont {G.~D.}\ \bibnamefont {Fuchs}},
  \bibinfo {author} {\bibfnamefont {L.~C.}\ \bibnamefont {Bassett}}, \ and\
  \bibinfo {author} {\bibfnamefont {D.~D.}\ \bibnamefont {Awschalom}},\ }\href
  {http://dx.doi.org/10.1126/science.1196436} {\bibfield  {journal} {\bibinfo
  {journal} {Science}\ }\textbf {\bibinfo {volume} {330}},\ \bibinfo {pages}
  {1212} (\bibinfo {year} {2010})}\BibitemShut {NoStop}%
\bibitem [{\citenamefont {Robledo}\ \emph {et~al.}(2011)\citenamefont
  {Robledo}, \citenamefont {Childress}, \citenamefont {Bernien}, \citenamefont
  {Hensen}, \citenamefont {Alkemade},\ and\ \citenamefont
  {Hanson}}]{Robledo:Nature2011}%
  \BibitemOpen
  \bibfield  {author} {\bibinfo {author} {\bibfnamefont {L.}~\bibnamefont
  {Robledo}}, \bibinfo {author} {\bibfnamefont {L.}~\bibnamefont {Childress}},
  \bibinfo {author} {\bibfnamefont {H.}~\bibnamefont {Bernien}}, \bibinfo
  {author} {\bibfnamefont {B.}~\bibnamefont {Hensen}}, \bibinfo {author}
  {\bibfnamefont {P.~F.~A.}\ \bibnamefont {Alkemade}}, \ and\ \bibinfo {author}
  {\bibfnamefont {R.}~\bibnamefont {Hanson}},\ }\href
  {http://dx.doi.org/10.1038/nature10401} {\bibfield  {journal} {\bibinfo
  {journal} {Nature}\ }\textbf {\bibinfo {volume} {477}},\ \bibinfo {pages}
  {574} (\bibinfo {year} {2011})}\BibitemShut {NoStop}%
\bibitem [{\citenamefont {Morton}\ \emph {et~al.}(2008)\citenamefont {Morton},
  \citenamefont {Tyryshkin}, \citenamefont {Brown}, \citenamefont {Shankar},
  \citenamefont {Lovett}, \citenamefont {Ardavan}, \citenamefont {Schenkel},
  \citenamefont {Haller}, \citenamefont {Ager},\ and\ \citenamefont
  {Lyon}}]{Morton2008}%
  \BibitemOpen
  \bibfield  {author} {\bibinfo {author} {\bibfnamefont {J.~J.~L.}\
  \bibnamefont {Morton}}, \bibinfo {author} {\bibfnamefont {A.~M.}\
  \bibnamefont {Tyryshkin}}, \bibinfo {author} {\bibfnamefont {R.~M.}\
  \bibnamefont {Brown}}, \bibinfo {author} {\bibfnamefont {S.}~\bibnamefont
  {Shankar}}, \bibinfo {author} {\bibfnamefont {B.~W.}\ \bibnamefont {Lovett}},
  \bibinfo {author} {\bibfnamefont {A.}~\bibnamefont {Ardavan}}, \bibinfo
  {author} {\bibfnamefont {T.}~\bibnamefont {Schenkel}}, \bibinfo {author}
  {\bibfnamefont {E.~E.}\ \bibnamefont {Haller}}, \bibinfo {author}
  {\bibfnamefont {J.~W.}\ \bibnamefont {Ager}}, \ and\ \bibinfo {author}
  {\bibfnamefont {S.~A.}\ \bibnamefont {Lyon}},\ }\href
  {http://dx.doi.org/10.1038/nature07295
  http://www.nature.com/nature/journal/v455/n7216/suppinfo/nature07295\_S1.html}
  {\bibfield  {journal} {\bibinfo  {journal} {Nature}\ }\textbf {\bibinfo
  {volume} {455}},\ \bibinfo {pages} {1085} (\bibinfo {year}
  {2008})}\BibitemShut {NoStop}%
\bibitem [{\citenamefont {Morello}\ \emph {et~al.}(2010)\citenamefont
  {Morello}, \citenamefont {Pla}, \citenamefont {Zwanenburg}, \citenamefont
  {Chan}, \citenamefont {Tan}, \citenamefont {Huebl}, \citenamefont {Mottonen},
  \citenamefont {Nugroho}, \citenamefont {Yang}, \citenamefont {van Donkelaar},
  \citenamefont {Alves}, \citenamefont {Jamieson}, \citenamefont {Escott},
  \citenamefont {Hollenberg}, \citenamefont {Clark},\ and\ \citenamefont
  {Dzurak}}]{Morello2010}%
  \BibitemOpen
  \bibfield  {author} {\bibinfo {author} {\bibfnamefont {A.}~\bibnamefont
  {Morello}}, \bibinfo {author} {\bibfnamefont {J.~J.}\ \bibnamefont {Pla}},
  \bibinfo {author} {\bibfnamefont {F.~A.}\ \bibnamefont {Zwanenburg}},
  \bibinfo {author} {\bibfnamefont {K.~W.}\ \bibnamefont {Chan}}, \bibinfo
  {author} {\bibfnamefont {K.~Y.}\ \bibnamefont {Tan}}, \bibinfo {author}
  {\bibfnamefont {H.}~\bibnamefont {Huebl}}, \bibinfo {author} {\bibfnamefont
  {M.}~\bibnamefont {Mottonen}}, \bibinfo {author} {\bibfnamefont {C.~D.}\
  \bibnamefont {Nugroho}}, \bibinfo {author} {\bibfnamefont {C.}~\bibnamefont
  {Yang}}, \bibinfo {author} {\bibfnamefont {J.~A.}\ \bibnamefont {van
  Donkelaar}}, \bibinfo {author} {\bibfnamefont {A.~D.~C.}\ \bibnamefont
  {Alves}}, \bibinfo {author} {\bibfnamefont {D.~N.}\ \bibnamefont {Jamieson}},
  \bibinfo {author} {\bibfnamefont {C.~C.}\ \bibnamefont {Escott}}, \bibinfo
  {author} {\bibfnamefont {L.~C.~L.}\ \bibnamefont {Hollenberg}}, \bibinfo
  {author} {\bibfnamefont {R.~G.}\ \bibnamefont {Clark}}, \ and\ \bibinfo
  {author} {\bibfnamefont {A.~S.}\ \bibnamefont {Dzurak}},\ }\href
  {http://dx.doi.org/10.1038/nature09392
  http://www.nature.com/nature/journal/v467/n7316/abs/nature09392.html\#supplementary-information}
  {\bibfield  {journal} {\bibinfo  {journal} {Nature}\ }\textbf {\bibinfo
  {volume} {467}},\ \bibinfo {pages} {687} (\bibinfo {year}
  {2010})}\BibitemShut {NoStop}%
\bibitem [{\citenamefont {Pla}\ \emph {et~al.}(2012)\citenamefont {Pla},
  \citenamefont {Tan}, \citenamefont {Dehollain}, \citenamefont {Lim},
  \citenamefont {Morton}, \citenamefont {Jamieson}, \citenamefont {Dzurak},\
  and\ \citenamefont {Morello}}]{Pla2012}%
  \BibitemOpen
  \bibfield  {author} {\bibinfo {author} {\bibfnamefont {J.~J.}\ \bibnamefont
  {Pla}}, \bibinfo {author} {\bibfnamefont {K.~Y.}\ \bibnamefont {Tan}},
  \bibinfo {author} {\bibfnamefont {J.~P.}\ \bibnamefont {Dehollain}}, \bibinfo
  {author} {\bibfnamefont {W.~H.}\ \bibnamefont {Lim}}, \bibinfo {author}
  {\bibfnamefont {J.~J.~L.}\ \bibnamefont {Morton}}, \bibinfo {author}
  {\bibfnamefont {D.~N.}\ \bibnamefont {Jamieson}}, \bibinfo {author}
  {\bibfnamefont {A.~S.}\ \bibnamefont {Dzurak}}, \ and\ \bibinfo {author}
  {\bibfnamefont {A.}~\bibnamefont {Morello}},\ }\href
  {http://dx.doi.org/10.1038/nature11449
  http://www.nature.com/nature/journal/v489/n7417/abs/nature11449.html\#supplementary-information}
  {\bibfield  {journal} {\bibinfo  {journal} {Nature}\ }\textbf {\bibinfo
  {volume} {489}},\ \bibinfo {pages} {541} (\bibinfo {year}
  {2012})}\BibitemShut {NoStop}%
\bibitem [{\citenamefont {Koehl}\ \emph {et~al.}(2011)\citenamefont {Koehl},
  \citenamefont {Buckley}, \citenamefont {Heremans}, \citenamefont {Calusine},\
  and\ \citenamefont {Awschalom}}]{Koehl11}%
  \BibitemOpen
  \bibfield  {author} {\bibinfo {author} {\bibfnamefont {W.~F.}\ \bibnamefont
  {Koehl}}, \bibinfo {author} {\bibfnamefont {B.~B.}\ \bibnamefont {Buckley}},
  \bibinfo {author} {\bibfnamefont {F.~J.}\ \bibnamefont {Heremans}}, \bibinfo
  {author} {\bibfnamefont {G.}~\bibnamefont {Calusine}}, \ and\ \bibinfo
  {author} {\bibfnamefont {D.~D.}\ \bibnamefont {Awschalom}},\ }\href@noop {}
  {\bibfield  {journal} {\bibinfo  {journal} {Nature}\ }\textbf {\bibinfo
  {volume} {479}},\ \bibinfo {pages} {84} (\bibinfo {year} {2011})}\BibitemShut
  {NoStop}%
\bibitem [{\citenamefont {Falk}\ \emph {et~al.}()\citenamefont {Falk},
  \citenamefont {Buckley}, \citenamefont {Calusine}, \citenamefont {Koehl},
  \citenamefont {Dobrovitski}, \citenamefont {Politi}, \citenamefont {Zorman},
  \citenamefont {Feng},\ and\ \citenamefont {Awschalom}}]{Falk2013}%
  \BibitemOpen
  \bibfield  {author} {\bibinfo {author} {\bibfnamefont {A.~L.}\ \bibnamefont
  {Falk}}, \bibinfo {author} {\bibfnamefont {B.~B.}\ \bibnamefont {Buckley}},
  \bibinfo {author} {\bibfnamefont {G.}~\bibnamefont {Calusine}}, \bibinfo
  {author} {\bibfnamefont {W.~F.}\ \bibnamefont {Koehl}}, \bibinfo {author}
  {\bibfnamefont {V.~V.}\ \bibnamefont {Dobrovitski}}, \bibinfo {author}
  {\bibfnamefont {A.}~\bibnamefont {Politi}}, \bibinfo {author} {\bibfnamefont
  {C.~A.}\ \bibnamefont {Zorman}}, \bibinfo {author} {\bibfnamefont {P.~X.-L.}\
  \bibnamefont {Feng}}, \ and\ \bibinfo {author} {\bibfnamefont {D.~D.}\
  \bibnamefont {Awschalom}},\ }\href@noop {} {\bibfield  {journal} {\bibinfo
  {journal} {Nature Commun.}\ }\textbf {\bibinfo {volume} {4}},\ \bibinfo
  {pages} {1819}}\BibitemShut {NoStop}%
\bibitem [{\citenamefont {Christle}\ \emph {et~al.}(2015)\citenamefont
  {Christle}, \citenamefont {Falk}, \citenamefont {Andrich}, \citenamefont
  {Klimov}, \citenamefont {Hassan}, \citenamefont {Son}, \citenamefont
  {Janz{\'e}n}, \citenamefont {Ohshima},\ and\ \citenamefont
  {Awschalom}}]{Christle2014}%
  \BibitemOpen
  \bibfield  {author} {\bibinfo {author} {\bibfnamefont {D.~J.}\ \bibnamefont
  {Christle}}, \bibinfo {author} {\bibfnamefont {A.~L.}\ \bibnamefont {Falk}},
  \bibinfo {author} {\bibfnamefont {P.}~\bibnamefont {Andrich}}, \bibinfo
  {author} {\bibfnamefont {P.~V.}\ \bibnamefont {Klimov}}, \bibinfo {author}
  {\bibfnamefont {J.~U.}\ \bibnamefont {Hassan}}, \bibinfo {author}
  {\bibfnamefont {N.~T.}\ \bibnamefont {Son}}, \bibinfo {author} {\bibfnamefont
  {E.}~\bibnamefont {Janz{\'e}n}}, \bibinfo {author} {\bibfnamefont
  {T.}~\bibnamefont {Ohshima}}, \ and\ \bibinfo {author} {\bibfnamefont
  {D.~D.}\ \bibnamefont {Awschalom}},\ }\href
  {http://dx.doi.org/10.1038/nmat4144} {\bibfield  {journal} {\bibinfo
  {journal} {Nat Mater}\ }\textbf {\bibinfo {volume} {14}},\ \bibinfo {pages}
  {160} (\bibinfo {year} {2015})}\BibitemShut {NoStop}%
\bibitem [{\citenamefont {Sipahigil}\ \emph {et~al.}(2014)\citenamefont
  {Sipahigil}, \citenamefont {Jahnke}, \citenamefont {Rogers}, \citenamefont
  {Teraji}, \citenamefont {Isoya}, \citenamefont {Zibrov}, \citenamefont
  {Jelezko},\ and\ \citenamefont {Lukin}}]{SipahigilPRL2014}%
  \BibitemOpen
  \bibfield  {author} {\bibinfo {author} {\bibfnamefont {A.}~\bibnamefont
  {Sipahigil}}, \bibinfo {author} {\bibfnamefont {K.~D.}\ \bibnamefont
  {Jahnke}}, \bibinfo {author} {\bibfnamefont {L.~J.}\ \bibnamefont {Rogers}},
  \bibinfo {author} {\bibfnamefont {T.}~\bibnamefont {Teraji}}, \bibinfo
  {author} {\bibfnamefont {J.}~\bibnamefont {Isoya}}, \bibinfo {author}
  {\bibfnamefont {A.~S.}\ \bibnamefont {Zibrov}}, \bibinfo {author}
  {\bibfnamefont {F.}~\bibnamefont {Jelezko}}, \ and\ \bibinfo {author}
  {\bibfnamefont {M.~D.}\ \bibnamefont {Lukin}},\ }\href {\doibase
  10.1103/PhysRevLett.113.113602} {\bibfield  {journal} {\bibinfo  {journal}
  {Phys. Rev. Lett.}\ }\textbf {\bibinfo {volume} {113}},\ \bibinfo {pages}
  {113602} (\bibinfo {year} {2014})}\BibitemShut {NoStop}%
\bibitem [{\citenamefont {Iwasaki}\ \emph {et~al.}(2015)\citenamefont
  {Iwasaki}, \citenamefont {Ishibashi}, \citenamefont {Miyamoto}, \citenamefont
  {Doi}, \citenamefont {Kobayashi}, \citenamefont {Miyazaki}, \citenamefont
  {Tahara}, \citenamefont {Jahnke}, \citenamefont {Rogers}, \citenamefont
  {Naydenov}, \citenamefont {Jelezko}, \citenamefont {Yamasaki}, \citenamefont
  {Nagamachi}, \citenamefont {Inubushi}, \citenamefont {Mizuochi},\ and\
  \citenamefont {Hatano}}]{Iwasaki2015}%
  \BibitemOpen
  \bibfield  {author} {\bibinfo {author} {\bibfnamefont {T.}~\bibnamefont
  {Iwasaki}}, \bibinfo {author} {\bibfnamefont {F.}~\bibnamefont {Ishibashi}},
  \bibinfo {author} {\bibfnamefont {Y.}~\bibnamefont {Miyamoto}}, \bibinfo
  {author} {\bibfnamefont {Y.}~\bibnamefont {Doi}}, \bibinfo {author}
  {\bibfnamefont {S.}~\bibnamefont {Kobayashi}}, \bibinfo {author}
  {\bibfnamefont {T.}~\bibnamefont {Miyazaki}}, \bibinfo {author}
  {\bibfnamefont {K.}~\bibnamefont {Tahara}}, \bibinfo {author} {\bibfnamefont
  {K.~D.}\ \bibnamefont {Jahnke}}, \bibinfo {author} {\bibfnamefont {L.~J.}\
  \bibnamefont {Rogers}}, \bibinfo {author} {\bibfnamefont {B.}~\bibnamefont
  {Naydenov}}, \bibinfo {author} {\bibfnamefont {F.}~\bibnamefont {Jelezko}},
  \bibinfo {author} {\bibfnamefont {S.}~\bibnamefont {Yamasaki}}, \bibinfo
  {author} {\bibfnamefont {S.}~\bibnamefont {Nagamachi}}, \bibinfo {author}
  {\bibfnamefont {T.}~\bibnamefont {Inubushi}}, \bibinfo {author}
  {\bibfnamefont {N.}~\bibnamefont {Mizuochi}}, \ and\ \bibinfo {author}
  {\bibfnamefont {M.}~\bibnamefont {Hatano}},\ }\href
  {http://dx.doi.org/10.1038/srep12882 10.1038/srep12882
  http://www.nature.com/articles/srep12882\#supplementary-information}
  {\bibfield  {journal} {\bibinfo  {journal} {Scientific Reports}\ }\textbf
  {\bibinfo {volume} {5}},\ \bibinfo {pages} {12882} (\bibinfo {year}
  {2015})}\BibitemShut {NoStop}%
\bibitem [{\citenamefont {Soltamov}\ \emph {et~al.}(2012)\citenamefont
  {Soltamov}, \citenamefont {Soltamova}, \citenamefont {Baranov},\ and\
  \citenamefont {Proskuryakov}}]{Soltamov12}%
  \BibitemOpen
  \bibfield  {author} {\bibinfo {author} {\bibfnamefont {V.~A.}\ \bibnamefont
  {Soltamov}}, \bibinfo {author} {\bibfnamefont {A.~A.}\ \bibnamefont
  {Soltamova}}, \bibinfo {author} {\bibfnamefont {P.~G.}\ \bibnamefont
  {Baranov}}, \ and\ \bibinfo {author} {\bibfnamefont {I.~I.}\ \bibnamefont
  {Proskuryakov}},\ }\href {\doibase 10.1103/PhysRevLett.108.226402} {\bibfield
   {journal} {\bibinfo  {journal} {Phys. Rev. Lett.}\ }\textbf {\bibinfo
  {volume} {108}},\ \bibinfo {pages} {226402} (\bibinfo {year}
  {2012})}\BibitemShut {NoStop}%
\bibitem [{\citenamefont {Kraus}\ \emph {et~al.}(2014)\citenamefont {Kraus},
  \citenamefont {Soltamov}, \citenamefont {Riedel}, \citenamefont {V\"ath},
  \citenamefont {Fuchs}, \citenamefont {Sperlich}, \citenamefont {Baranov},
  \citenamefont {Dyakonov},\ and\ \citenamefont {Astakhov}}]{NatPhys14}%
  \BibitemOpen
  \bibfield  {author} {\bibinfo {author} {\bibfnamefont {H.}~\bibnamefont
  {Kraus}}, \bibinfo {author} {\bibfnamefont {V.~A.}\ \bibnamefont {Soltamov}},
  \bibinfo {author} {\bibfnamefont {D.}~\bibnamefont {Riedel}}, \bibinfo
  {author} {\bibfnamefont {S.}~\bibnamefont {V\"ath}}, \bibinfo {author}
  {\bibfnamefont {F.}~\bibnamefont {Fuchs}}, \bibinfo {author} {\bibfnamefont
  {A.}~\bibnamefont {Sperlich}}, \bibinfo {author} {\bibfnamefont {P.~G.}\
  \bibnamefont {Baranov}}, \bibinfo {author} {\bibfnamefont {V.}~\bibnamefont
  {Dyakonov}}, \ and\ \bibinfo {author} {\bibfnamefont {G.~V.}\ \bibnamefont
  {Astakhov}},\ }\href {\doibase 10.1038/nphys2826} {\bibfield  {journal}
  {\bibinfo  {journal} {Nature Physics}\ }\textbf {\bibinfo {volume} {10}},\
  \bibinfo {pages} {157} (\bibinfo {year} {2014})}\BibitemShut {NoStop}%
\bibitem [{\citenamefont {Sz\'asz}\ \emph {et~al.}(2015)\citenamefont
  {Sz\'asz}, \citenamefont {Iv\'ady}, \citenamefont {Abrikosov}, \citenamefont
  {Janz\'en}, \citenamefont {Bockstedte},\ and\ \citenamefont
  {Gali}}]{Szasz2015}%
  \BibitemOpen
  \bibfield  {author} {\bibinfo {author} {\bibfnamefont {K.}~\bibnamefont
  {Sz\'asz}}, \bibinfo {author} {\bibfnamefont {V.}~\bibnamefont {Iv\'ady}},
  \bibinfo {author} {\bibfnamefont {I.~A.}\ \bibnamefont {Abrikosov}}, \bibinfo
  {author} {\bibfnamefont {E.}~\bibnamefont {Janz\'en}}, \bibinfo {author}
  {\bibfnamefont {M.}~\bibnamefont {Bockstedte}}, \ and\ \bibinfo {author}
  {\bibfnamefont {A.}~\bibnamefont {Gali}},\ }\href {\doibase
  10.1103/PhysRevB.91.121201} {\bibfield  {journal} {\bibinfo  {journal} {Phys.
  Rev. B}\ }\textbf {\bibinfo {volume} {91}},\ \bibinfo {pages} {121201}
  (\bibinfo {year} {2015})}\BibitemShut {NoStop}%
\bibitem [{\citenamefont {Kolesov}\ \emph {et~al.}(2013)\citenamefont
  {Kolesov}, \citenamefont {Xia}, \citenamefont {Reuter}, \citenamefont
  {Jamali}, \citenamefont {St\"ohr}, \citenamefont {Inal}, \citenamefont
  {Siyushev},\ and\ \citenamefont {Wrachtrup}}]{KolesovPRL2013}%
  \BibitemOpen
  \bibfield  {author} {\bibinfo {author} {\bibfnamefont {R.}~\bibnamefont
  {Kolesov}}, \bibinfo {author} {\bibfnamefont {K.}~\bibnamefont {Xia}},
  \bibinfo {author} {\bibfnamefont {R.}~\bibnamefont {Reuter}}, \bibinfo
  {author} {\bibfnamefont {M.}~\bibnamefont {Jamali}}, \bibinfo {author}
  {\bibfnamefont {R.}~\bibnamefont {St\"ohr}}, \bibinfo {author} {\bibfnamefont
  {T.}~\bibnamefont {Inal}}, \bibinfo {author} {\bibfnamefont {P.}~\bibnamefont
  {Siyushev}}, \ and\ \bibinfo {author} {\bibfnamefont {J.}~\bibnamefont
  {Wrachtrup}},\ }\href {\doibase 10.1103/PhysRevLett.111.120502} {\bibfield
  {journal} {\bibinfo  {journal} {Phys. Rev. Lett.}\ }\textbf {\bibinfo
  {volume} {111}},\ \bibinfo {pages} {120502} (\bibinfo {year}
  {2013})}\BibitemShut {NoStop}%
\bibitem [{\citenamefont {Longdell}\ \emph {et~al.}(2004)\citenamefont
  {Longdell}, \citenamefont {Sellars},\ and\ \citenamefont
  {Manson}}]{LongdellPRL2014}%
  \BibitemOpen
  \bibfield  {author} {\bibinfo {author} {\bibfnamefont {J.~J.}\ \bibnamefont
  {Longdell}}, \bibinfo {author} {\bibfnamefont {M.~J.}\ \bibnamefont
  {Sellars}}, \ and\ \bibinfo {author} {\bibfnamefont {N.~B.}\ \bibnamefont
  {Manson}},\ }\href {\doibase 10.1103/PhysRevLett.93.130503} {\bibfield
  {journal} {\bibinfo  {journal} {Phys. Rev. Lett.}\ }\textbf {\bibinfo
  {volume} {93}},\ \bibinfo {pages} {130503} (\bibinfo {year}
  {2004})}\BibitemShut {NoStop}%
\bibitem [{\citenamefont {Clausen}\ \emph {et~al.}(2011)\citenamefont
  {Clausen}, \citenamefont {Usmani}, \citenamefont {Bussieres}, \citenamefont
  {Sangouard}, \citenamefont {Afzelius}, \citenamefont {de~Riedmatten},\ and\
  \citenamefont {Gisin}}]{Clausen2011}%
  \BibitemOpen
  \bibfield  {author} {\bibinfo {author} {\bibfnamefont {C.}~\bibnamefont
  {Clausen}}, \bibinfo {author} {\bibfnamefont {I.}~\bibnamefont {Usmani}},
  \bibinfo {author} {\bibfnamefont {F.}~\bibnamefont {Bussieres}}, \bibinfo
  {author} {\bibfnamefont {N.}~\bibnamefont {Sangouard}}, \bibinfo {author}
  {\bibfnamefont {M.}~\bibnamefont {Afzelius}}, \bibinfo {author}
  {\bibfnamefont {H.}~\bibnamefont {de~Riedmatten}}, \ and\ \bibinfo {author}
  {\bibfnamefont {N.}~\bibnamefont {Gisin}},\ }\href
  {http://dx.doi.org/10.1038/nature09662
  http://www.nature.com/nature/journal/v469/n7331/abs/10.1038-nature09662-unlocked.html\#supplementary-information}
  {\bibfield  {journal} {\bibinfo  {journal} {Nature}\ }\textbf {\bibinfo
  {volume} {469}},\ \bibinfo {pages} {508} (\bibinfo {year}
  {2011})}\BibitemShut {NoStop}%
\bibitem [{\citenamefont {de~Riedmatten}\ \emph {et~al.}(2008)\citenamefont
  {de~Riedmatten}, \citenamefont {Afzelius}, \citenamefont {Staudt},
  \citenamefont {Simon},\ and\ \citenamefont {Gisin}}]{DeRiedmatten2008}%
  \BibitemOpen
  \bibfield  {author} {\bibinfo {author} {\bibfnamefont {H.}~\bibnamefont
  {de~Riedmatten}}, \bibinfo {author} {\bibfnamefont {M.}~\bibnamefont
  {Afzelius}}, \bibinfo {author} {\bibfnamefont {M.~U.}\ \bibnamefont
  {Staudt}}, \bibinfo {author} {\bibfnamefont {C.}~\bibnamefont {Simon}}, \
  and\ \bibinfo {author} {\bibfnamefont {N.}~\bibnamefont {Gisin}},\ }\href
  {http://dx.doi.org/10.1038/nature07607
  http://www.nature.com/nature/journal/v456/n7223/suppinfo/nature07607\_S1.html}
  {\bibfield  {journal} {\bibinfo  {journal} {Nature}\ }\textbf {\bibinfo
  {volume} {456}},\ \bibinfo {pages} {773} (\bibinfo {year}
  {2008})}\BibitemShut {NoStop}%
\bibitem [{\citenamefont {Seo}\ \emph {et~al.}(2016)\citenamefont {Seo},
  \citenamefont {Govoni},\ and\ \citenamefont {Galli}}]{Seo2016}%
  \BibitemOpen
  \bibfield  {author} {\bibinfo {author} {\bibfnamefont {H.}~\bibnamefont
  {Seo}}, \bibinfo {author} {\bibfnamefont {M.}~\bibnamefont {Govoni}}, \ and\
  \bibinfo {author} {\bibfnamefont {G.}~\bibnamefont {Galli}},\ }\href
  {http://dx.doi.org/10.1038/srep20803 10.1038/srep20803
  http://www.nature.com/articles/srep20803\#supplementary-information}
  {\bibfield  {journal} {\bibinfo  {journal} {Scientific Reports}\ }\textbf
  {\bibinfo {volume} {6}},\ \bibinfo {pages} {20803} (\bibinfo {year}
  {2016})}\BibitemShut {NoStop}%
\bibitem [{\citenamefont {Janz{\'{e}}n}\ and\ \citenamefont
  {Magnusson}(2005)}]{magnusson2005}%
  \BibitemOpen
  \bibfield  {author} {\bibinfo {author} {\bibfnamefont {E.}~\bibnamefont
  {Janz{\'{e}}n}}\ and\ \bibinfo {author} {\bibfnamefont {B.}~\bibnamefont
  {Magnusson}},\ }in\ \href {\doibase
  10.4028/www.scientific.net/MSF.483-485.341} {\emph {\bibinfo {booktitle}
  {Silicon Carbide and Related Materials 2004}}},\ \bibinfo {series} {Materials
  Science Forum}, Vol.\ \bibinfo {volume} {483}\ (\bibinfo  {publisher} {Trans
  Tech Publications},\ \bibinfo {year} {2005})\ pp.\ \bibinfo {pages}
  {341--346}\BibitemShut {NoStop}%
\bibitem [{\citenamefont {Gordon}\ \emph {et~al.}(2015)\citenamefont {Gordon},
  \citenamefont {Janotti},\ and\ \citenamefont {Van~de Walle}}]{Gordon2015}%
  \BibitemOpen
  \bibfield  {author} {\bibinfo {author} {\bibfnamefont {L.}~\bibnamefont
  {Gordon}}, \bibinfo {author} {\bibfnamefont {A.}~\bibnamefont {Janotti}}, \
  and\ \bibinfo {author} {\bibfnamefont {C.~G.}\ \bibnamefont {Van~de Walle}},\
  }\href {\doibase 10.1103/PhysRevB.92.045208} {\bibfield  {journal} {\bibinfo
  {journal} {Phys. Rev. B}\ }\textbf {\bibinfo {volume} {92}},\ \bibinfo
  {pages} {045208} (\bibinfo {year} {2015})}\BibitemShut {NoStop}%
\bibitem [{\citenamefont {Falk}\ \emph {et~al.}(2014)\citenamefont {Falk},
  \citenamefont {Klimov}, \citenamefont {Buckley}, \citenamefont {Iv\'ady},
  \citenamefont {Abrikosov}, \citenamefont {Calusine}, \citenamefont {Koehl},
  \citenamefont {Gali},\ and\ \citenamefont {Awschalom}}]{Falk2014}%
  \BibitemOpen
  \bibfield  {author} {\bibinfo {author} {\bibfnamefont {A.~L.}\ \bibnamefont
  {Falk}}, \bibinfo {author} {\bibfnamefont {P.~V.}\ \bibnamefont {Klimov}},
  \bibinfo {author} {\bibfnamefont {B.~B.}\ \bibnamefont {Buckley}}, \bibinfo
  {author} {\bibfnamefont {V.}~\bibnamefont {Iv\'ady}}, \bibinfo {author}
  {\bibfnamefont {I.~A.}\ \bibnamefont {Abrikosov}}, \bibinfo {author}
  {\bibfnamefont {G.}~\bibnamefont {Calusine}}, \bibinfo {author}
  {\bibfnamefont {W.~F.}\ \bibnamefont {Koehl}}, \bibinfo {author}
  {\bibfnamefont {A.}~\bibnamefont {Gali}}, \ and\ \bibinfo {author}
  {\bibfnamefont {D.~D.}\ \bibnamefont {Awschalom}},\ }\href {\doibase
  10.1103/PhysRevLett.112.187601} {\bibfield  {journal} {\bibinfo  {journal}
  {Phys. Rev. Lett.}\ }\textbf {\bibinfo {volume} {112}},\ \bibinfo {pages}
  {187601} (\bibinfo {year} {2014})}\BibitemShut {NoStop}%
\bibitem [{\citenamefont {Kohan}\ \emph {et~al.}(2000)\citenamefont {Kohan},
  \citenamefont {Ceder}, \citenamefont {Morgan},\ and\ \citenamefont {Van~de
  Walle}}]{Kohan2000}%
  \BibitemOpen
  \bibfield  {author} {\bibinfo {author} {\bibfnamefont {A.~F.}\ \bibnamefont
  {Kohan}}, \bibinfo {author} {\bibfnamefont {G.}~\bibnamefont {Ceder}},
  \bibinfo {author} {\bibfnamefont {D.}~\bibnamefont {Morgan}}, \ and\ \bibinfo
  {author} {\bibfnamefont {C.~G.}\ \bibnamefont {Van~de Walle}},\ }\href
  {\doibase 10.1103/PhysRevB.61.15019} {\bibfield  {journal} {\bibinfo
  {journal} {Phys. Rev. B}\ }\textbf {\bibinfo {volume} {61}},\ \bibinfo
  {pages} {15019} (\bibinfo {year} {2000})}\BibitemShut {NoStop}%
\bibitem [{\citenamefont {Janotti}\ and\ \citenamefont {Van~de
  Walle}(2007)}]{Janotti2007}%
  \BibitemOpen
  \bibfield  {author} {\bibinfo {author} {\bibfnamefont {A.}~\bibnamefont
  {Janotti}}\ and\ \bibinfo {author} {\bibfnamefont {C.~G.}\ \bibnamefont
  {Van~de Walle}},\ }\href {\doibase 10.1103/PhysRevB.76.165202} {\bibfield
  {journal} {\bibinfo  {journal} {Phys. Rev. B}\ }\textbf {\bibinfo {volume}
  {76}},\ \bibinfo {pages} {165202} (\bibinfo {year} {2007})}\BibitemShut
  {NoStop}%
\bibitem [{\citenamefont {Lany}\ and\ \citenamefont
  {Zunger}(2008)}]{LanyZunger08}%
  \BibitemOpen
  \bibfield  {author} {\bibinfo {author} {\bibfnamefont {S.}~\bibnamefont
  {Lany}}\ and\ \bibinfo {author} {\bibfnamefont {A.}~\bibnamefont {Zunger}},\
  }\href {\doibase 10.1103/PhysRevB.78.235104} {\bibfield  {journal} {\bibinfo
  {journal} {Phys. Rev. B}\ }\textbf {\bibinfo {volume} {78}},\ \bibinfo
  {pages} {235104} (\bibinfo {year} {2008})}\BibitemShut {NoStop}%
\bibitem [{\citenamefont {De\'ak}\ \emph {et~al.}(2010)\citenamefont {De\'ak},
  \citenamefont {Aradi}, \citenamefont {Frauenheim}, \citenamefont {Janz\'en},\
  and\ \citenamefont {Gali}}]{Deak2010}%
  \BibitemOpen
  \bibfield  {author} {\bibinfo {author} {\bibfnamefont {P.}~\bibnamefont
  {De\'ak}}, \bibinfo {author} {\bibfnamefont {B.}~\bibnamefont {Aradi}},
  \bibinfo {author} {\bibfnamefont {T.}~\bibnamefont {Frauenheim}}, \bibinfo
  {author} {\bibfnamefont {E.}~\bibnamefont {Janz\'en}}, \ and\ \bibinfo
  {author} {\bibfnamefont {A.}~\bibnamefont {Gali}},\ }\href {\doibase
  10.1103/PhysRevB.81.153203} {\bibfield  {journal} {\bibinfo  {journal} {Phys.
  Rev. B}\ }\textbf {\bibinfo {volume} {81}},\ \bibinfo {pages} {153203}
  (\bibinfo {year} {2010})}\BibitemShut {NoStop}%
\bibitem [{\citenamefont {Freysoldt}\ \emph {et~al.}(2014)\citenamefont
  {Freysoldt}, \citenamefont {Grabowski}, \citenamefont {Hickel}, \citenamefont
  {Neugebauer}, \citenamefont {Kresse}, \citenamefont {Janotti},\ and\
  \citenamefont {Van~de Walle}}]{FreysoldtRMP2014}%
  \BibitemOpen
  \bibfield  {author} {\bibinfo {author} {\bibfnamefont {C.}~\bibnamefont
  {Freysoldt}}, \bibinfo {author} {\bibfnamefont {B.}~\bibnamefont
  {Grabowski}}, \bibinfo {author} {\bibfnamefont {T.}~\bibnamefont {Hickel}},
  \bibinfo {author} {\bibfnamefont {J.}~\bibnamefont {Neugebauer}}, \bibinfo
  {author} {\bibfnamefont {G.}~\bibnamefont {Kresse}}, \bibinfo {author}
  {\bibfnamefont {A.}~\bibnamefont {Janotti}}, \ and\ \bibinfo {author}
  {\bibfnamefont {C.~G.}\ \bibnamefont {Van~de Walle}},\ }\href {\doibase
  10.1103/RevModPhys.86.253} {\bibfield  {journal} {\bibinfo  {journal} {Rev.
  Mod. Phys.}\ }\textbf {\bibinfo {volume} {86}},\ \bibinfo {pages} {253}
  (\bibinfo {year} {2014})}\BibitemShut {NoStop}%
\bibitem [{\citenamefont {Ceder}\ and\ \citenamefont
  {Persson}(2013)}]{ceder2013supercomputers}%
  \BibitemOpen
  \bibfield  {author} {\bibinfo {author} {\bibfnamefont {G.}~\bibnamefont
  {Ceder}}\ and\ \bibinfo {author} {\bibfnamefont {K.}~\bibnamefont
  {Persson}},\ }\href@noop {} {\bibfield  {journal} {\bibinfo  {journal}
  {Scientific American (Dec 2013)}\ } (\bibinfo {year} {2013})}\BibitemShut
  {NoStop}%
\bibitem [{\citenamefont {Curtarolo}\ \emph {et~al.}(2013)\citenamefont
  {Curtarolo}, \citenamefont {Hart}, \citenamefont {Nardelli}, \citenamefont
  {Mingo}, \citenamefont {Sanvito},\ and\ \citenamefont
  {Levy}}]{curtarolo2013high}%
  \BibitemOpen
  \bibfield  {author} {\bibinfo {author} {\bibfnamefont {S.}~\bibnamefont
  {Curtarolo}}, \bibinfo {author} {\bibfnamefont {G.~L.}\ \bibnamefont {Hart}},
  \bibinfo {author} {\bibfnamefont {M.~B.}\ \bibnamefont {Nardelli}}, \bibinfo
  {author} {\bibfnamefont {N.}~\bibnamefont {Mingo}}, \bibinfo {author}
  {\bibfnamefont {S.}~\bibnamefont {Sanvito}}, \ and\ \bibinfo {author}
  {\bibfnamefont {O.}~\bibnamefont {Levy}},\ }\href@noop {} {\bibfield
  {journal} {\bibinfo  {journal} {Nature materials}\ }\textbf {\bibinfo
  {volume} {12}},\ \bibinfo {pages} {191} (\bibinfo {year} {2013})}\BibitemShut
  {NoStop}%
\bibitem [{\citenamefont {Gali}\ \emph {et~al.}(2010)\citenamefont {Gali},
  \citenamefont {G\"allstr\"om}, \citenamefont {Son},\ and\ \citenamefont
  {Janz\'en}}]{Gali10}%
  \BibitemOpen
  \bibfield  {author} {\bibinfo {author} {\bibfnamefont {A.}~\bibnamefont
  {Gali}}, \bibinfo {author} {\bibfnamefont {A.}~\bibnamefont {G\"allstr\"om}},
  \bibinfo {author} {\bibfnamefont {N.}~\bibnamefont {Son}}, \ and\ \bibinfo
  {author} {\bibfnamefont {E.}~\bibnamefont {Janz\'en}},\ }\href@noop {}
  {\bibfield  {journal} {\bibinfo  {journal} {Mater. Sci. Forum}\ }\textbf
  {\bibinfo {volume} {645-648}},\ \bibinfo {pages} {395} (\bibinfo {year}
  {2010})}\BibitemShut {NoStop}%
\bibitem [{\citenamefont {Doherty}\ \emph {et~al.}(2013)\citenamefont
  {Doherty}, \citenamefont {Manson}, \citenamefont {Delaney}, \citenamefont
  {Jelezko}, \citenamefont {Wrachtrup},\ and\ \citenamefont
  {Hollenberg}}]{Doherty2013}%
  \BibitemOpen
  \bibfield  {author} {\bibinfo {author} {\bibfnamefont {M.~W.}\ \bibnamefont
  {Doherty}}, \bibinfo {author} {\bibfnamefont {N.~B.}\ \bibnamefont {Manson}},
  \bibinfo {author} {\bibfnamefont {P.}~\bibnamefont {Delaney}}, \bibinfo
  {author} {\bibfnamefont {F.}~\bibnamefont {Jelezko}}, \bibinfo {author}
  {\bibfnamefont {J.}~\bibnamefont {Wrachtrup}}, \ and\ \bibinfo {author}
  {\bibfnamefont {L.~C.}\ \bibnamefont {Hollenberg}},\ }\href {\doibase
  http://dx.doi.org/10.1016/j.physrep.2013.02.001} {\bibfield  {journal}
  {\bibinfo  {journal} {Physics Reports}\ }\textbf {\bibinfo {volume} {528}},\
  \bibinfo {pages} {1 } (\bibinfo {year} {2013})},\ \bibinfo {note} {the
  nitrogen-vacancy colour centre in diamond}\BibitemShut {NoStop}%
\bibitem [{\citenamefont {Maze}\ \emph {et~al.}(2010)\citenamefont {Maze},
  \citenamefont {Gali}, \citenamefont {Togan}, \citenamefont {Chu},
  \citenamefont {Trifonov}, \citenamefont {Kaxiras},\ and\ \citenamefont
  {Lukin}}]{Maze:2010}%
  \BibitemOpen
  \bibfield  {author} {\bibinfo {author} {\bibfnamefont {J.~R.}\ \bibnamefont
  {Maze}}, \bibinfo {author} {\bibfnamefont {A.}~\bibnamefont {Gali}}, \bibinfo
  {author} {\bibfnamefont {E.}~\bibnamefont {Togan}}, \bibinfo {author}
  {\bibfnamefont {Y.}~\bibnamefont {Chu}}, \bibinfo {author} {\bibfnamefont
  {A.}~\bibnamefont {Trifonov}}, \bibinfo {author} {\bibfnamefont
  {E.}~\bibnamefont {Kaxiras}}, \ and\ \bibinfo {author} {\bibfnamefont
  {M.~D.}\ \bibnamefont {Lukin}},\ }\href@noop {} {\bibfield  {journal}
  {\bibinfo  {journal} {e-print arXiv:}\ }\textbf {\bibinfo {volume}
  {quant-ph/1010.1338}} (\bibinfo {year} {2010})}\BibitemShut {NoStop}%
\bibitem [{\citenamefont {Doherty}\ \emph {et~al.}(2011)\citenamefont
  {Doherty}, \citenamefont {Manson}, \citenamefont {Delaney},\ and\
  \citenamefont {Hollenberg}}]{Doherty2011}%
  \BibitemOpen
  \bibfield  {author} {\bibinfo {author} {\bibfnamefont {M.~W.}\ \bibnamefont
  {Doherty}}, \bibinfo {author} {\bibfnamefont {N.~B.}\ \bibnamefont {Manson}},
  \bibinfo {author} {\bibfnamefont {P.}~\bibnamefont {Delaney}}, \ and\
  \bibinfo {author} {\bibfnamefont {L.~C.~L.}\ \bibnamefont {Hollenberg}},\
  }\href {http://stacks.iop.org/1367-2630/13/i=2/a=025019} {\bibfield
  {journal} {\bibinfo  {journal} {New Journal of Physics}\ }\textbf {\bibinfo
  {volume} {13}},\ \bibinfo {pages} {025019} (\bibinfo {year}
  {2011})}\BibitemShut {NoStop}%
\bibitem [{\citenamefont {Baranov}\ \emph {et~al.}(2005)\citenamefont
  {Baranov}, \citenamefont {Il'in}, \citenamefont {Mokhov}, \citenamefont
  {Muzafarova}, \citenamefont {Orlinskii},\ and\ \citenamefont
  {Schmidt}}]{Baranov2005}%
  \BibitemOpen
  \bibfield  {author} {\bibinfo {author} {\bibfnamefont {P.}~\bibnamefont
  {Baranov}}, \bibinfo {author} {\bibfnamefont {I.}~\bibnamefont {Il'in}},
  \bibinfo {author} {\bibfnamefont {E.}~\bibnamefont {Mokhov}}, \bibinfo
  {author} {\bibfnamefont {M.}~\bibnamefont {Muzafarova}}, \bibinfo {author}
  {\bibfnamefont {S.}~\bibnamefont {Orlinskii}}, \ and\ \bibinfo {author}
  {\bibfnamefont {J.}~\bibnamefont {Schmidt}},\ }\href {\doibase
  10.1134/1.2142873} {\bibfield  {journal} {\bibinfo  {journal} {JETP Letters}\
  }\textbf {\bibinfo {volume} {82}},\ \bibinfo {pages} {441} (\bibinfo {year}
  {2005})}\BibitemShut {NoStop}%
\bibitem [{\citenamefont {Son}\ \emph {et~al.}(2006)\citenamefont {Son},
  \citenamefont {Carlsson}, \citenamefont {ul~Hassan}, \citenamefont
  {Janz\'en}, \citenamefont {Umeda}, \citenamefont {Isoya}, \citenamefont
  {Gali}, \citenamefont {Bockstedte}, \citenamefont {Morishita}, \citenamefont
  {Ohshima},\ and\ \citenamefont {Itoh}}]{Son2006}%
  \BibitemOpen
  \bibfield  {author} {\bibinfo {author} {\bibfnamefont {N.}~\bibnamefont
  {Son}}, \bibinfo {author} {\bibfnamefont {P.}~\bibnamefont {Carlsson}},
  \bibinfo {author} {\bibfnamefont {J.}~\bibnamefont {ul~Hassan}}, \bibinfo
  {author} {\bibfnamefont {E.}~\bibnamefont {Janz\'en}}, \bibinfo {author}
  {\bibfnamefont {T.}~\bibnamefont {Umeda}}, \bibinfo {author} {\bibfnamefont
  {J.}~\bibnamefont {Isoya}}, \bibinfo {author} {\bibfnamefont
  {A.}~\bibnamefont {Gali}}, \bibinfo {author} {\bibfnamefont {M.}~\bibnamefont
  {Bockstedte}}, \bibinfo {author} {\bibfnamefont {N.}~\bibnamefont
  {Morishita}}, \bibinfo {author} {\bibfnamefont {T.}~\bibnamefont {Ohshima}},
  \ and\ \bibinfo {author} {\bibfnamefont {H.}~\bibnamefont {Itoh}},\ }\href
  {\doibase 10.1103/PhysRevLett.96.055501} {\bibfield  {journal} {\bibinfo
  {journal} {Phys. Rev. Lett.}\ }\textbf {\bibinfo {volume} {96}},\ \bibinfo
  {pages} {055501} (\bibinfo {year} {2006})}\BibitemShut {NoStop}%
\bibitem [{\citenamefont {Hohenberg}\ and\ \citenamefont
  {Kohn}(1964)}]{Hohenberg64}%
  \BibitemOpen
  \bibfield  {author} {\bibinfo {author} {\bibfnamefont {P.}~\bibnamefont
  {Hohenberg}}\ and\ \bibinfo {author} {\bibfnamefont {W.}~\bibnamefont
  {Kohn}},\ }\href {\doibase 10.1103/PhysRev.136.B864} {\bibfield  {journal}
  {\bibinfo  {journal} {Phys. Rev.}\ }\textbf {\bibinfo {volume} {136}},\
  \bibinfo {pages} {B864} (\bibinfo {year} {1964})}\BibitemShut {NoStop}%
\bibitem [{\citenamefont {Kohn}\ and\ \citenamefont {Sham}(1965)}]{Kohn65}%
  \BibitemOpen
  \bibfield  {author} {\bibinfo {author} {\bibfnamefont {W.}~\bibnamefont
  {Kohn}}\ and\ \bibinfo {author} {\bibfnamefont {L.~J.}\ \bibnamefont
  {Sham}},\ }\href {\doibase 10.1103/PhysRev.140.A1133} {\bibfield  {journal}
  {\bibinfo  {journal} {Phys. Rev.}\ }\textbf {\bibinfo {volume} {140}},\
  \bibinfo {pages} {A1133} (\bibinfo {year} {1965})}\BibitemShut {NoStop}%
\bibitem [{\citenamefont {Gali}\ \emph {et~al.}(2009)\citenamefont {Gali},
  \citenamefont {Janz\'en}, \citenamefont {De\'ak}, \citenamefont {Kresse},\
  and\ \citenamefont {Kaxiras}}]{Gali:PRL2009}%
  \BibitemOpen
  \bibfield  {author} {\bibinfo {author} {\bibfnamefont {A.}~\bibnamefont
  {Gali}}, \bibinfo {author} {\bibfnamefont {E.}~\bibnamefont {Janz\'en}},
  \bibinfo {author} {\bibfnamefont {P.}~\bibnamefont {De\'ak}}, \bibinfo
  {author} {\bibfnamefont {G.}~\bibnamefont {Kresse}}, \ and\ \bibinfo {author}
  {\bibfnamefont {E.}~\bibnamefont {Kaxiras}},\ }\href@noop {} {\bibfield
  {journal} {\bibinfo  {journal} {Phys. Rev. Lett.}\ }\textbf {\bibinfo
  {volume} {103}},\ \bibinfo {pages} {186404} (\bibinfo {year}
  {2009})}\BibitemShut {NoStop}%
\bibitem [{\citenamefont {Perdew}\ \emph {et~al.}(1996)\citenamefont {Perdew},
  \citenamefont {Burke},\ and\ \citenamefont {Ernzerhof}}]{PBE}%
  \BibitemOpen
  \bibfield  {author} {\bibinfo {author} {\bibfnamefont {J.~P.}\ \bibnamefont
  {Perdew}}, \bibinfo {author} {\bibfnamefont {K.}~\bibnamefont {Burke}}, \
  and\ \bibinfo {author} {\bibfnamefont {M.}~\bibnamefont {Ernzerhof}},\ }\href
  {\doibase 10.1103/PhysRevLett.77.3865} {\bibfield  {journal} {\bibinfo
  {journal} {Phys. Rev. Lett.}\ }\textbf {\bibinfo {volume} {77}},\ \bibinfo
  {pages} {3865} (\bibinfo {year} {1996})}\BibitemShut {NoStop}%
\bibitem [{\citenamefont {Armiento}\ and\ \citenamefont
  {Mattsson}(2005)}]{AM05}%
  \BibitemOpen
  \bibfield  {author} {\bibinfo {author} {\bibfnamefont {R.}~\bibnamefont
  {Armiento}}\ and\ \bibinfo {author} {\bibfnamefont {A.~E.}\ \bibnamefont
  {Mattsson}},\ }\href {\doibase 10.1103/PhysRevB.72.085108} {\bibfield
  {journal} {\bibinfo  {journal} {Phys. Rev. B}\ }\textbf {\bibinfo {volume}
  {72}},\ \bibinfo {pages} {085108} (\bibinfo {year} {2005})}\BibitemShut
  {NoStop}%
\bibitem [{\citenamefont {Heyd}\ \emph {et~al.}(2003)\citenamefont {Heyd},
  \citenamefont {Scuseria},\ and\ \citenamefont {Ernzerhof}}]{HSE03}%
  \BibitemOpen
  \bibfield  {author} {\bibinfo {author} {\bibfnamefont {J.}~\bibnamefont
  {Heyd}}, \bibinfo {author} {\bibfnamefont {G.~E.}\ \bibnamefont {Scuseria}},
  \ and\ \bibinfo {author} {\bibfnamefont {M.}~\bibnamefont {Ernzerhof}},\
  }\href {\doibase 10.1063/1.1564060} {\bibfield  {journal} {\bibinfo
  {journal} {J. Chem. Phys.}\ }\textbf {\bibinfo {volume} {118}},\ \bibinfo
  {pages} {8207} (\bibinfo {year} {2003})}\BibitemShut {NoStop}%
\bibitem [{\citenamefont {Heyd}\ \emph {et~al.}(2006)\citenamefont {Heyd},
  \citenamefont {Scuseria},\ and\ \citenamefont {Ernzerhof}}]{HSE06}%
  \BibitemOpen
  \bibfield  {author} {\bibinfo {author} {\bibfnamefont {J.}~\bibnamefont
  {Heyd}}, \bibinfo {author} {\bibfnamefont {G.~E.}\ \bibnamefont {Scuseria}},
  \ and\ \bibinfo {author} {\bibfnamefont {M.}~\bibnamefont {Ernzerhof}},\
  }\href {\doibase 10.1063/1.2204597} {\bibfield  {journal} {\bibinfo
  {journal} {J. Chem. Phys.}\ }\textbf {\bibinfo {volume} {124}},\ \bibinfo
  {pages} {219906} (\bibinfo {year} {2006})}\BibitemShut {NoStop}%
\bibitem [{\citenamefont {Heyd}\ \emph {et~al.}(2005)\citenamefont {Heyd},
  \citenamefont {Peralta}, \citenamefont {Scuseria},\ and\ \citenamefont
  {Martin}}]{HSE05}%
  \BibitemOpen
  \bibfield  {author} {\bibinfo {author} {\bibfnamefont {J.}~\bibnamefont
  {Heyd}}, \bibinfo {author} {\bibfnamefont {J.~E.}\ \bibnamefont {Peralta}},
  \bibinfo {author} {\bibfnamefont {G.~E.}\ \bibnamefont {Scuseria}}, \ and\
  \bibinfo {author} {\bibfnamefont {R.~L.}\ \bibnamefont {Martin}},\ }\href
  {\doibase 10.1063/1.2085170} {\bibfield  {journal} {\bibinfo  {journal} {The
  Journal of Chemical Physics}\ }\textbf {\bibinfo {volume} {123}},\ \bibinfo
  {pages} {174101} (\bibinfo {year} {2005})},\ \Eprint
  {http://arxiv.org/abs/http://dx.doi.org/10.1063/1.2085170}
  {http://dx.doi.org/10.1063/1.2085170} \BibitemShut {NoStop}%
\bibitem [{\citenamefont {Sz\'asz}\ \emph {et~al.}(2013)\citenamefont
  {Sz\'asz}, \citenamefont {Hornos}, \citenamefont {Marsman},\ and\
  \citenamefont {Gali}}]{Szasz2013}%
  \BibitemOpen
  \bibfield  {author} {\bibinfo {author} {\bibfnamefont {K.}~\bibnamefont
  {Sz\'asz}}, \bibinfo {author} {\bibfnamefont {T.}~\bibnamefont {Hornos}},
  \bibinfo {author} {\bibfnamefont {M.}~\bibnamefont {Marsman}}, \ and\
  \bibinfo {author} {\bibfnamefont {A.}~\bibnamefont {Gali}},\ }\href {\doibase
  10.1103/PhysRevB.88.075202} {\bibfield  {journal} {\bibinfo  {journal} {Phys.
  Rev. B}\ }\textbf {\bibinfo {volume} {88}},\ \bibinfo {pages} {075202}
  (\bibinfo {year} {2013})}\BibitemShut {NoStop}%
\bibitem [{\citenamefont {Kresse}\ and\ \citenamefont {Hafner}(1994)}]{VASP}%
  \BibitemOpen
  \bibfield  {author} {\bibinfo {author} {\bibfnamefont {G.}~\bibnamefont
  {Kresse}}\ and\ \bibinfo {author} {\bibfnamefont {J.}~\bibnamefont
  {Hafner}},\ }\href {\doibase 10.1103/PhysRevB.49.14251} {\bibfield  {journal}
  {\bibinfo  {journal} {Phys. Rev. B}\ }\textbf {\bibinfo {volume} {49}},\
  \bibinfo {pages} {14251} (\bibinfo {year} {1994})}\BibitemShut {NoStop}%
\bibitem [{\citenamefont {Kresse}\ and\ \citenamefont
  {Furthm\"uller}(1996)}]{VASP2}%
  \BibitemOpen
  \bibfield  {author} {\bibinfo {author} {\bibfnamefont {G.}~\bibnamefont
  {Kresse}}\ and\ \bibinfo {author} {\bibfnamefont {J.}~\bibnamefont
  {Furthm\"uller}},\ }\href {\doibase 10.1103/PhysRevB.54.11169} {\bibfield
  {journal} {\bibinfo  {journal} {Phys. Rev. B}\ }\textbf {\bibinfo {volume}
  {54}},\ \bibinfo {pages} {11169} (\bibinfo {year} {1996})}\BibitemShut
  {NoStop}%
\bibitem [{\citenamefont {Bl\"ochl}(1994)}]{PAW}%
  \BibitemOpen
  \bibfield  {author} {\bibinfo {author} {\bibfnamefont {P.~E.}\ \bibnamefont
  {Bl\"ochl}},\ }\href {\doibase 10.1103/PhysRevB.50.17953} {\bibfield
  {journal} {\bibinfo  {journal} {Phys. Rev. B}\ }\textbf {\bibinfo {volume}
  {50}},\ \bibinfo {pages} {17953} (\bibinfo {year} {1994})}\BibitemShut
  {NoStop}%
\bibitem [{\citenamefont {Kresse}\ and\ \citenamefont
  {Joubert}(1999)}]{Kresse99}%
  \BibitemOpen
  \bibfield  {author} {\bibinfo {author} {\bibfnamefont {G.}~\bibnamefont
  {Kresse}}\ and\ \bibinfo {author} {\bibfnamefont {D.}~\bibnamefont
  {Joubert}},\ }\href {\doibase 10.1103/PhysRevB.59.1758} {\bibfield  {journal}
  {\bibinfo  {journal} {Phys. Rev. B}\ }\textbf {\bibinfo {volume} {59}},\
  \bibinfo {pages} {1758} (\bibinfo {year} {1999})}\BibitemShut {NoStop}%
\bibitem [{\citenamefont {Monkhorst}\ and\ \citenamefont
  {Pack}(1976)}]{monkhorst1976special}%
  \BibitemOpen
  \bibfield  {author} {\bibinfo {author} {\bibfnamefont {H.~J.}\ \bibnamefont
  {Monkhorst}}\ and\ \bibinfo {author} {\bibfnamefont {J.~D.}\ \bibnamefont
  {Pack}},\ }\href@noop {} {\bibfield  {journal} {\bibinfo  {journal} {Physical
  review B}\ }\textbf {\bibinfo {volume} {13}},\ \bibinfo {pages} {5188}
  (\bibinfo {year} {1976})}\BibitemShut {NoStop}%
\bibitem [{\citenamefont {Iv\'ady}\ \emph {et~al.}(2014)\citenamefont
  {Iv\'ady}, \citenamefont {Simon}, \citenamefont {Maze}, \citenamefont
  {Abrikosov},\ and\ \citenamefont {Gali}}]{Ivady2014}%
  \BibitemOpen
  \bibfield  {author} {\bibinfo {author} {\bibfnamefont {V.}~\bibnamefont
  {Iv\'ady}}, \bibinfo {author} {\bibfnamefont {T.}~\bibnamefont {Simon}},
  \bibinfo {author} {\bibfnamefont {J.~R.}\ \bibnamefont {Maze}}, \bibinfo
  {author} {\bibfnamefont {I.~A.}\ \bibnamefont {Abrikosov}}, \ and\ \bibinfo
  {author} {\bibfnamefont {A.}~\bibnamefont {Gali}},\ }\href@noop {} {\bibfield
   {journal} {\bibinfo  {journal} {Phys. Rev. B}\ }\textbf {\bibinfo {volume}
  {90}},\ \bibinfo {pages} {235205} (\bibinfo {year} {2014})}\BibitemShut
  {NoStop}%
\bibitem [{\citenamefont {Doherty}\ \emph {et~al.}(2014)\citenamefont
  {Doherty}, \citenamefont {Struzhkin}, \citenamefont {Simpson}, \citenamefont
  {McGuinness}, \citenamefont {Meng}, \citenamefont {Stacey}, \citenamefont
  {Karle}, \citenamefont {Hemley}, \citenamefont {Manson}, \citenamefont
  {Hollenberg},\ and\ \citenamefont {Prawer}}]{Doherty2014}%
  \BibitemOpen
  \bibfield  {author} {\bibinfo {author} {\bibfnamefont {M.~W.}\ \bibnamefont
  {Doherty}}, \bibinfo {author} {\bibfnamefont {V.~V.}\ \bibnamefont
  {Struzhkin}}, \bibinfo {author} {\bibfnamefont {D.~A.}\ \bibnamefont
  {Simpson}}, \bibinfo {author} {\bibfnamefont {L.~P.}\ \bibnamefont
  {McGuinness}}, \bibinfo {author} {\bibfnamefont {Y.}~\bibnamefont {Meng}},
  \bibinfo {author} {\bibfnamefont {A.}~\bibnamefont {Stacey}}, \bibinfo
  {author} {\bibfnamefont {T.~J.}\ \bibnamefont {Karle}}, \bibinfo {author}
  {\bibfnamefont {R.~J.}\ \bibnamefont {Hemley}}, \bibinfo {author}
  {\bibfnamefont {N.~B.}\ \bibnamefont {Manson}}, \bibinfo {author}
  {\bibfnamefont {L.~C.~L.}\ \bibnamefont {Hollenberg}}, \ and\ \bibinfo
  {author} {\bibfnamefont {S.}~\bibnamefont {Prawer}},\ }\href {\doibase
  10.1103/PhysRevLett.112.047601} {\bibfield  {journal} {\bibinfo  {journal}
  {Phys. Rev. Lett.}\ }\textbf {\bibinfo {volume} {112}},\ \bibinfo {pages}
  {047601} (\bibinfo {year} {2014})}\BibitemShut {NoStop}%
\bibitem [{\citenamefont {Falk}\ \emph {et~al.}(2015)\citenamefont {Falk},
  \citenamefont {Klimov}, \citenamefont {Iv\'ady}, \citenamefont {Sz\'asz},
  \citenamefont {Christle}, \citenamefont {Koehl}, \citenamefont {Gali},\ and\
  \citenamefont {Awschalom}}]{FalkPRL2015}%
  \BibitemOpen
  \bibfield  {author} {\bibinfo {author} {\bibfnamefont {A.~L.}\ \bibnamefont
  {Falk}}, \bibinfo {author} {\bibfnamefont {P.~V.}\ \bibnamefont {Klimov}},
  \bibinfo {author} {\bibfnamefont {V.}~\bibnamefont {Iv\'ady}}, \bibinfo
  {author} {\bibfnamefont {K.}~\bibnamefont {Sz\'asz}}, \bibinfo {author}
  {\bibfnamefont {D.~J.}\ \bibnamefont {Christle}}, \bibinfo {author}
  {\bibfnamefont {W.~F.}\ \bibnamefont {Koehl}}, \bibinfo {author}
  {\bibfnamefont {A.}~\bibnamefont {Gali}}, \ and\ \bibinfo {author}
  {\bibfnamefont {D.~D.}\ \bibnamefont {Awschalom}},\ }\href {\doibase
  10.1103/PhysRevLett.114.247603} {\bibfield  {journal} {\bibinfo  {journal}
  {Phys. Rev. Lett.}\ }\textbf {\bibinfo {volume} {114}},\ \bibinfo {pages}
  {247603} (\bibinfo {year} {2015})}\BibitemShut {NoStop}%
\end{thebibliography}%

\end{document}